\documentclass[aps, pra, showpacs, twocolumn, amsmath, amssymb]{revtex4}

\usepackage{dcolumn} 
\usepackage{bm} 
\usepackage{graphicx}
\usepackage{float}
\usepackage[english]{babel}

\newcommand{\bfsfG}{\mbox{\sffamily\bfseries{G}}}
\newcommand{\bfsfI}{\mbox{\sffamily\bfseries{I}}}

\newcommand{\bfsfT}{\mbox{\sffamily\bfseries{T}}}
\newcommand{\bfsfV}{\mbox{\sffamily\bfseries{V}}}

\begin{document}

\title{F\"orster resonance energy transfer rate and local density of optical states are uncorrelated in any dielectric nanophotonic medium}

\author{Martijn Wubs}
\email{mwubs@fotonik.dtu.dk}
\affiliation{Department of Photonics Engineering, Technical University of Denmark, DK-2800 Kgs. Lyngby, Denmark}
\affiliation{Center for Nanostructured Graphene, Technical University of Denmark, DK-2800 Kongens Lyngby, Denmark}

\author{Willem L. Vos}
\email{w.l.vos@utwente.nl, world wide web: www.photonicbandgaps.com}
\affiliation{Complex Photonic Systems (COPS), MESA+ Institute for Nanotechnology, University of Twente, P.O. Box 217, 7500 AE Enschede, The Netherlands}

\date{July 22nd, 2015}

\begin{abstract}
Motivated by the ongoing debate about nanophotonic control of F{\"o}rster resonance energy transfer (FRET), notably by the local density of optical states (LDOS), we study an analytic model system wherein a pair of ideal dipole emitters - donor and acceptor - exhibit energy transfer in the vicinity of an ideal mirror.
The FRET rate is controlled by the mirror up to distances comparable to the donor-acceptor distance, that is, the few-nanometer range.
For vanishing distance, we find a complete inhibition or a four-fold enhancement, depending on dipole orientation.
For mirror distances on the wavelength scale, where the well-known `Drexhage' modification of the spontaneous-emission rate occurs, the FRET rate is constant.
Hence there is no correlation between the F{\"o}rster (or total) energy transfer rate and the LDOS.
At any distance to the mirror, the total energy transfer between a closely-spaced donor and acceptor is dominated by F{\"o}rster transfer, \emph{i.e.}, by the static dipole-dipole interaction that yields the characteristic inverse-sixth-power donor-acceptor distance in homogeneous media.
Generalizing to arbitrary inhomogeneous media with weak dispersion and weak absorption in the frequency overlap range of donor and acceptor, we derive two main theoretical results.
Firstly, the spatial dependence of the F{\"o}rster energy transfer rate does not depend on frequency, hence not on the LDOS.
Secondly the FRET rate is expressed as a frequency integral of the imaginary part of the Green function.
This leads to an approximate FRET rate in terms of the LDOS integrated over a huge bandwidth from zero frequency to about $10 \times$ the donor emission frequency, corresponding to the vacuum-ultraviolet.
Even then, the broadband LDOS hardly contributes to the energy transfer rates.
Using our analytical expressions, we plot transfer rates at an experimentally relevant emission wavelength $\lambda = 628$ nm that reveal nm-ranged distances, and discuss practical consequences including quantum information processing.

\end{abstract}

\maketitle

\section{Introduction}

A well-known optical interaction between pairs of quantum emitters - such as excited atoms, ions, molecules, or quantum dots - is F{\"o}rster resonance energy transfer (FRET).
In this process, first identified in a seminal 1948 paper by F{\"o}rster, one quantum of excitation energy is transferred from a first emitter, called a donor, to a second emitter that is referred to as an acceptor~\cite{Forster1948AP}.
FRET is the dominant energy transfer mechanism between emitters in nanometer proximity, since the rate has a characteristic $(r_{F}/r_{\rm da})^6$ distance dependence (with $r_{F}$ the F{\"o}rster radius and $r_{\rm da}$ the distance between donor and acceptor).
Other means to control a FRET system are traditionally the spectral properties of the coupled emitters - the overlap between the donor's emission spectrum and the acceptor's absorptions spectrum - or the relative orientations of the dipole moments~\cite{Forster1948AP, Lakowicz2006}.
FRET plays a central role in the photosynthetic apparatus of plants and bacteria~\cite{Grondelle1994BBAB, Scholes2003ARPC}.
Many applications are based on FRET, ranging from photovoltaics~\cite{Chanyawadee2009PRL, Buhbut2010AN},  lighting~\cite{Baldo2000N, Vohra2010AN}, and magneto-optics~\cite{Vincent2011PRB}, to sensing~\cite{Medintz2003NM} where molecular distances~\cite{Stryer1978ARB, Schuler2002N}, and interactions are probed~\cite{Garcia2004CPC, Carriba2008NM}.
FRET is also relevant to the manipulation, storage, and transfer of quantum information~\cite{John1991PRB, Barenco1995PRL, Lovett2003PRB, Reina2004PRL, Nazir:2005a, Unold2005PRL}.

Modern nanofabrication techniques have stimulated the relevant question whether F{\"o}rster transfer can be controlled purely by means of the nanophotonic environment, while leaving the FRET pair geometrically and chemically unchanged.
In many situations, the effect of the nanophotonic environment can be expressed in terms of the local density of optical states (LDOS) that counts the number of photon modes available for emission, and is interpreted as the density of vacuum fluctuations~\cite{Sprik1996EPL, Barnes1998JMO}.
An assumption behind many recent FRET studies has therefore been that {\em if} there is an effect of the nanophotonic environment on FRET rates, then it should be possible to find a general law that describes the functional dependence of FRET rates on the LDOS.
While it is an assumption, it is a fruitful one as it allows for experimental verification.
Curiously, different dependencies of FRET rates on the LDOS were reported in a number of experimental studies, leading to the debate how the F{\"o}rster energy transfer rate depends on the LDOS.
Pioneering work by Andrew and Barnes indicated that the transfer rate depends linearly on the donor decay rate and thus the LDOS at the donor emission frequency~\cite{Andrew2000S}, as was confirmed elsewhere~\cite{Nakamura2005PRB}.
The linear relation between the two rates found in~\cite{Andrew2000S} was supported by the theory of Ref.~\onlinecite{Dung2002PRA}, but only fortuitously within a limited parameter regime.
Subsequent experiments suggested a transfer rate independent of the LDOS~\cite{Dood2005PRB}, a dependence on the LDOS squared~\cite{Nakamura2006PRB}, or qualitative effects~\cite{Kolaric2007CM, Yang2008OL}.
Possible reasons for the disparity in these observations include insufficient control on the donor-acceptor distance, on incomplete pairing of every donor to only one acceptor, or on cross-talk between neighboring donor-acceptor pairs.

Recently, the relation between F{\"o}rster transfer and the LDOS was studied using precisely-defined, isolated, and efficient donor-acceptor pairs~\cite{Blum2012PRL}.
The distance between donor and acceptor molecules was fixed by covalently binding them to the opposite ends of a 15 basepair long double-stranded DNA.
A precise control over the LDOS was realized by positioning the donor-acceptor pairs at well-defined distances to a metallic mirror~\cite{Drexhage1970JL, Barnes1998JMO, Chance1978ACP}.
The outcome of this experimental study was that the F{\"o}rster transfer rate is \emph{independent} on the optical LDOS, as was confirmed by theoretical considerations.
Consequently, the F{\"o}rster transfer efficiency is greatest for a vanishing LDOS, hence in a 3D photonic band gap crystal~\cite{Leistikow2011PRL}.
Similar results were obtained with different light sources (rare-earth ions), and with different cavities~\cite{Rabouw2014NC, Konrad2015NS}.
On the other hand, a linear relation between LDOS and FRET rate was reported in experiments with donors and acceptors at a few nanometers from metal surfaces~\cite{Ghenuche2014NL, Torres:2015a}, while Ref.~\cite{GonzagaGaleana:2013a} reported no general theoretical relationship between LDOS and FRET rate near a metallic sphere.
In Ref.~\onlinecite{Tumkur2014FD} the measured dependence of the FRET rate on the LDOS was reported to be weak for single FRET pairs; an observed drop of the total energy transfer rate of a donor close to a surface was mainly attributed to the simple fact that fewer statistically distributed acceptors are available close to the surface; recent theoretical work on collective energy transfer supports these results in the dilute limit~\cite{Poddubny:2015a}.

Most theoretical papers agree that both the energy transfer rate and the spontaneous-emission rate can be expressed in terms of the Green function of the nanophotonic medium.
One may argue that energy transfer and optical LDOS are therefore related.
But one may also argue to the contrary, since the energy transfer rate depends on the total Green function describing propagation from donor to acceptor, whereas the spontaneous-emission rate depends on the imaginary part of the Green function at the donor position only~\cite{Dung2002PRA}.
It is not clear whether this situation entails correlations between the two quantities and if so, what is their functional relationship.
Therefore, and in view of the different results in the literature, we decided that a study of a simple analytical model is timely.

In this paper, we first study energy transfer in a prototypical nanophotonic medium, namely the Drexhage geometry~\cite{Drexhage1970JL}, near an ideal mirror.
This is one of the simplest inhomogeneous dielectric media for which position-dependent spontaneous-emission rates are analytically known~\cite{Chance1978ACP}.
Here we show that the spatial dependence of both the total and the F{\"o}rster resonant energy transfer rates can be calculated  analytically.
Such exact results have a value of their own, and allow for a critical and straightforward assessment of possible correlations between the FRET rate and the LDOS.
After studying the phenomenology near an ideal mirror, we derive general results for {\em arbitrary inhomogeneous} weakly dispersive  media, and thereby find F{\" o}rster transfer rates that generalize the well-known $1/(n^4 r_{\rm da}^6)$ dependence of the homogeneous nondispersive medium with refractive index $n$.
Specifically, using the energy-transfer theory by Dung, Kn{\" o}ll, and Welsch~\cite{Dung2002PRA} and the Green-function properties derived in Ref.~\cite{Wubs:2004a} as starting points, we derive a simple and important new consequence: for the large class of photonic media that have little dispersion and absorption in the donor-acceptor frequency overlap range, there is no correlation between the LDOS and the position-dependent FRET rate.
Nevertheless, the FRET rate is controlled by the distance to the mirror, but only at mirror distances comparable to or smaller than the donor-acceptor separation.

\begin{figure}[t]
\includegraphics[width=1.0\columnwidth]{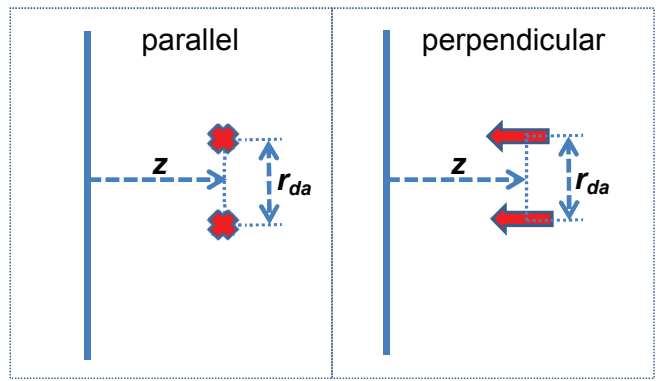}
\caption{(Color online)
We study pairs of donor and acceptor dipoles that are separated by a distance $r_{\rm da}$, and located at a distance $z$ from an ideal mirror.
We focus on two configurations, both with dipoles oriented perpendicular to the position difference vector of donor and acceptor $({\bf \hat \mu}_{\rm d}, {\bf \hat \mu}_{\rm a}) \perp ({\bf r}_{\rm d} - {\bf r}_{\rm a})$:
(a) Both dipole moments of donor and acceptor are parallel to the mirror surface (`parallel configuration', $\parallel$) and parallel to each other; (b) Both dipole moments of donor and acceptor are perpendicular to the mirror surface (`perpendicular configuration', $\perp$) and parallel to each other.
}
\label{Fig:situation}
\end{figure}

\section{Physical processes and geometry}
We study energy transfer from a single donor to a single acceptor separated by a distance $r_{\rm da} = |{\bf r}_{\rm a} - {\bf r}_{\rm d}|$.
To limit parameter space, we focus on situations in which the donor and the acceptor have the same distance $z$ to the mirror, and where the dipole moments of dipole and acceptor point in the same direction.
In the parallel ($\parallel$) configuration shown in Figure~\ref{Fig:situation}(a), both dipole moments are oriented parallel to the mirror, and the dipoles point normal to the mirror in the perpendicular ($\perp$) configuration of Figure~\ref{Fig:situation}(b).

The total energy transfer rate $\gamma_{\rm da}$ between a donor and an acceptor dipole
in any nanophotonic environment is given by
%
\begin{equation}\label{energy-transfer-rate}
\gamma_{\rm da} = \int_{-\infty}^{\infty} \mbox{d}\omega\,\sigma_{\rm a}(\omega)\, w({\bf r}_{\rm a},{\bf r}_{\rm d},\omega)\, \sigma_{\rm d}(\omega),
\end{equation}
%
where $\sigma_{\rm d,a}(\omega)$ are the donor (single-photon) emission and acceptor (single-photon) absorption spectra in free space~\cite{May:2000a, Dung2002PRA}.
All effects of the nanophotonic environment are contained in the transfer amplitude squared $w({\bf r}_{\rm a},{\bf r}_{\rm d},\omega)$ that can be expressed in terms of the Green function ${\bfsfG}({\bf r}_{\rm a},{\bf r}_{\rm d},\omega)$ of the medium, and the donor and acceptor dipole moments ${\bm \mu}_{\rm d}, {\bm \mu}_{\rm a}$ respectively, as
%
\begin{equation}\label{wamplitude}
w({\bf r}_{\rm a},{\bf r}_{\rm d},\omega) = \frac{2\pi}{\hbar^{2}} \left(\frac{\omega^{2}}{\varepsilon_{0} c^{2}}\right)^{2}  \; |{\bm  \mu}_{\rm a}^{*}\cdot {\bfsfG}({\bf r}_{\rm a},{\bf r}_{\rm d},\omega)\cdot {\bm  \mu}_{\rm d}|^2.
\end{equation}
%
These expressions for the total energy transfer rate were originally derived in an important paper by Dung, Kn{\" o}ll, and Welsch for a general class of nanophotonic media that may exhibit both frequency-dispersion and absorption~\cite{Dung2002PRA} \footnote{Our function $w$ is the same as $\tilde w$ in Ref.~\cite{Dung2002PRA}, Eq.~(44).}.
The total energy transfer rate is the combined effect of both radiative and F{\"o}rster energy transfer processes.
We will first study total energy transfer rates, and in Sec.~\ref{Sec:Foerster_versus_total} we discuss what fraction of this energy transfer is F{\"o}rster resonance energy transfer.

For the energy transfer rate Eq.~(\ref{energy-transfer-rate}) we only need to know the Green function in the frequency interval where the donor and acceptor spectra overlap appreciably.
In case of molecules that have among the broadest bandwidths, this overlap has a typical relative bandwidth of only a few percent.
Hence it is reasonable to neglect absorption and material dispersion in this narrow overlap region.
Thus $\varepsilon({\bf r},\omega)$ can be approximated by a real-valued frequency-independent dielectric function $\varepsilon({\bf r})$.
The corresponding Green function ${\bfsfG}({\bf r},{\bf r}',\omega)$ is the solution of the usual wave equation for light
%
\begin{equation}\label{eqforG}
-\nabla\times\nabla \times {\bfsfG}({\bf r},{\bf r}',\omega) + \varepsilon({\bf r}) \left(\frac{\omega}{c}\right)^{2} {\bfsfG}({\bf r},{\bf r}',\omega) =
\delta({\bf r} - {\bf r}') {\bfsfI},
\end{equation}
%
with a localized source on the right-hand side \footnote{Our definition of the Green function agrees with  Refs.~\cite{DeVries:1998a,Wubs:2004a} and differs by a minus sign from Refs.~\cite{Dung2002PRA,Novotny2012}.}.
Unlike $\varepsilon({\bf r})$, the Green function ${\bfsfG}({\bf r},{\bf r}',\omega)$ is frequency-dependent and complex-valued.

While the energy transfer rate in Eq.~(\ref{energy-transfer-rate}) evidently depends on the donor and acceptor spectra $\sigma_{\rm d}(\omega)$ and $\sigma_{\rm a}(\omega)$, we are in this paper more interested in the dependence on the environment as given in Eq.~(\ref{wamplitude}).
We therefore assume that the donor and acceptor overlap in a narrow-frequency region in which the transfer amplitude $w({\bf r}_{\rm a},{\bf r}_{\rm d},\omega)$ varies negligibly with frequency, so we can approximate the energy transfer rate by
\begin{equation}\label{energy-transfer-rate-approx}
\bar \gamma_{\rm da} = w({\bf r}_{\rm a},{\bf r}_{\rm d},\omega_{\rm da}) \int_{-\infty}^{\infty} \mbox{d}\omega\,\sigma_{\rm a}(\omega)\,  \sigma_{\rm d}(\omega),
\end{equation}
where $\omega_{\rm da}$ is the frequency where the integrand in the overlap integral assumes its maximal value.
The overlap integral is the same for all nanophotonic environments, so that the ratio of energy transfer rates in two different environments only depends on the ratio of $w({\bf r}_{\rm a},{\bf r}_{\rm d},\omega_{\rm da})$ in both environments.

Spontaneous emission of the donor is a process that competes with the energy transfer to the acceptor.
The donor spontaneous-emission rate $\gamma_{\rm se}({\bf r},\Omega)$ at position ${\bf r}$ with real-valued dipole moment ${\bm \mu} = \mu \hat{\bm \mu}$ and transition frequency $\omega_{\rm d}$ is expressed in terms of the imaginary part of the Green function of the medium as
%
\begin{equation}\label{gammadby2}
\gamma_{\rm se}({\bf r}_{\rm d},\omega_{\rm d}) =  - \left(\frac{2 \omega_{\rm d}^{2}}{\hbar\varepsilon_{0} c^{2}}\right){\bm \mu}\cdot{\rm Im}[\bfsfG({\bf r}_{\rm d},{\bf r}_{\rm d},\omega_{\rm d})]\cdot{\bm \mu}
\end{equation}
%
or $\gamma({\bf r}_{\rm d},\omega_{\rm d},{\bm \mu}) = \pi\mu^{2}\omega_{\rm d}\rho_{\rm p}({\bf r}_{\rm d},\omega_{\rm d},\hat{\bm \mu})/(3 \hbar \varepsilon_{0})$ in terms of the partial LDOS $\rho_{\rm p}({\bf r}_{\rm d},\omega_{\rm d},\hat{\bm \mu}) = - (6 \omega_{\rm d}/\pi c^{2}) \hat{\bm \mu} \cdot {\rm Im}[\bfsfG({\bf r}_{\rm d},{\bf r}_{\rm d},\omega_{\rm d})]\cdot \hat{\bm \mu}$, where $\hat{\bm \mu}$ is a dipole-orientation unit vector~\cite{Sprik1996EPL, Novotny2012}.
The optical density of states (LDOS) is then defined as the dipole-orientation-averaged partial LDOS~\cite{Novotny2012}.
In general both the LDOS and the partial LDOS for any dipole orientation are fixed once the partial LDOS is known for nine independent dipole orientations, but for planar systems considered here, the two directions $\perp$ and $\parallel$ suffice for a complete description~\cite{Vos:2009a} \footnote{What is for convenience called LDOS should in the following be understood to be the partial LDOS.}.
We do not average over dipole orientations, as we are interested in possible correlations between energy transfer and spontaneous-emission rates for a fixed dipole orientation.
%
\begin{table}[t]
\begin{tabular}{ |l|l| }
\hline
  \multicolumn{2}{|c|}{Glossary of transfer and emission rates} \\
  \hline
  $\gamma_{\rm da}$ & total donor-acceptor energy transfer rate, Eq.~(\ref{energy-transfer-rate}) \\
  $\bar\gamma_{\rm da}$ & narrowband approximation of transfer rate, Eq.~(\ref{energy-transfer-rate-approx}) \\
  $\gamma_{\rm se}$ & spontaneous emission rate of the donor,  Eq.~(\ref{gammadby2}) \\
  $\gamma_{\rm F}$ & exact FRET rate from donor to acceptor, Eq.~(\ref{eq:FRETrate2}) \\
  $\gamma_{\rm F}^{\rm (L)}$ & broadband LDOS approximated FRET rate, Eq.~(\ref{eq:FRETrateLDOS})  \\
  $\tilde\gamma_{\rm F}^{(\rm HF)}$ & high-frequency approximated FRET rate, Eq.~(\ref{eq:FRETrateHF}) \\
  \hline
\end{tabular}
\caption{Symbols for the various energy transfer and emission rates used in this paper, with their defining equations.}
\label{Table1}
\end{table}

To proceed we need to compute Green functions.
First, the Green tensor in a homogeneous medium with real-valued refractive index $n$ is given by~\cite{DeVries:1998a}
%
\begin{eqnarray}\label{Ghom_realspace}
&&\bfsfG_{\rm h}({\bf r}_{1},{\bf r}_{2},\omega)  =  \bfsfG_{\rm h}({\bf r},\omega)  = \nonumber\\
& & - \frac{e^{w}}{4 \pi r}\left[P(w)\bfsfI + Q(w)\hat{\bf r}\otimes \hat{\bf r}\right] +
\frac{1}{3 (n\omega/c)^2} \delta({\bf r}) \bfsfI,
\end{eqnarray}
%
where ${\bf r} = {\bf r}_{1}-{\bf r}_{2}$, the functions $P,Q$ are defined as $P(w) \equiv (1-w^{-1} + w^{-2})$ and $Q(w) \equiv (-1 + 3 w^{-1} - 3 w^{-2}) $, and the argument equals $w = (i n \omega r/c)$.
For $n=1$, $\bfsfG_{\rm h}$ equals the free-space Green function, denoted by $\bfsfG_{0}$.
For distances much smaller than an optical wavelength $(r = |{\bf r}| \ll \lambda = 2 \pi c/(n\omega))$, the Green function scales as $\bfsfG_{\rm h}({\bf r},\omega) \propto 1/(n^{2}r^{3})$.
From Eqs.~(\ref{energy-transfer-rate}) and (\ref{wamplitude}) we then obtain the characteristic scaling of the F{\" o}rster transfer rate as $\gamma_{\rm da} \propto 1/(n^{4} r_{\rm da}^{6})$: the F{\" o}rster transfer rate strongly \emph{decreases} with  increasing donor-acceptor distance and with increasing refractive index.
In contrast, it follows from Eq.~(\ref{gammadby2}) that the spontaneous-emission rate $\gamma_{\rm se}$ in a homogeneous medium is \emph{enhanced} by a factor $n$ compared to free space.
More refined analyses that include local-field effects likewise predict a spontaneous-emission enhancement~\cite{Schuurmans1998PRL}.
These major differences between the energy transfer rate and the emission rate in a homogeneous medium already give an inkling on the behavior in nanophotonics.

Next, we determine the Green function of an ideal flat mirror within an otherwise homogeneous medium with refractive index $n$.
While the function can be found with various methods~\cite{Chance1978ACP, Matloob:2000a}, we briefly show how it is obtained by generalizing the multiple-scattering formalism of Ref.~\onlinecite{Wubs2004PRE} for infinitely thin planes.
In the usual mixed Fourier-real-space representation $({\bf k}_{\parallel},z)$ relevant to planar systems with  translational invariance in the $(x,y)$-directions, the homogeneous-medium Green function $\bfsfG_{\rm h}({\bf k}_{\parallel}, z, z',\omega)$ becomes
%
\begin{equation}\label{Ghplane}
\bfsfG_{\rm h} = \left(\begin{array}{ccc} 1 & 0 & 0 \\
                                     0 & k_{z}^{2} & - k_{\parallel}k_{z}s_{zz'} \\
                                     0 & - k_{\parallel}k_{z}s_{zz'} & k_{\parallel}^{2}                                     \end{array}\right)\frac{c^{2}}{(n\omega)^{2}} g_{\rm h} + \frac{\delta(z-z')}{(n\omega/c)^{2}}\hat {\bf z} \hat {\bf z},
\end{equation}
%
where the scalar Green function is given by $g_{\rm h} = g_{\rm h}({\bf k}_{\parallel}, z, z',\omega) = \exp(2 i k_{z}|z - z'|)/(2 i k_{z})$, $k_{z} = (n\omega^{2}/c^{2} - k_{\parallel}^{2})^{1/2}$, $s_{zz'} = sign(z-z')$ and the matrix is represented in the basis $(\hat {\bf s}_{\bf k}, \hat {\bf p}_{\bf k},\hat {\bf z})$, where ${\bf k}$ is the wave vector of the incoming light, $\hat {\bf z}$ is the positive-z-direction, $\hat {\bf s}_{\bf k}$ is the direction of s-polarized light (out of the plane of incidence), and $\hat {\bf p}_{\bf k}$ points perpendicular to $\hat {\bf z}$ in the plane of incidence.
An infinitely thin plane at $z=0$ that scatters light can be described by a T-matrix $\bfsfT({\bf k}_{\parallel},\omega)$, in terms of which the Green function becomes
%
\begin{equation}\label{GintermsofT}
\bfsfG(z,z') = \bfsfG_{\rm h}(z,z') + \bfsfG_{\rm h}(z,0) \bfsfT\, \bfsfG_{\rm h}(0,z'),
\end{equation}
%
where the $({\bf k}_{\parallel},\omega)$ dependence was dropped.
It was found in Ref.~\onlinecite{Wubs2004PRE} that for an infinitely thin plane that models a finite-thickness dielectric slab of dielectric constant $\varepsilon$, the T-matrix assumes a diagonal form in the same basis as $\bfsfG_{\rm h}$ in Eq.~(\ref{Ghplane}), in particular $\bfsfT = diag(T^{\rm ss},T^{\rm pp}, 0)$.
The infinitely thin plane becomes a perfectly reflecting mirror if we choose for example a lossless Drude response with $\varepsilon = 1 - \omega_{\rm p}^{2}/\omega^{2}$, in the limit of an infinite plasma frequency $\omega_{\rm p}\rightarrow \infty$.
Hence the T-matrix for a perfect mirror in a homogeneous dielectric has nonzero diagonal components $T^{\rm ss} = - 2 i k_{z}$ and $T^{\rm pp} = - 2 i (n\omega/c)^{2}/k_{z}$.
The ideal mirror divides space into two optically disconnected half spaces, and below we only consider the half space $z \ge 0$.
It then follows that the Green function for the ideal mirror is written in terms of homogeneous-medium Green functions as
\begin{equation}\label{Gmirrorassumof3terms}
\bfsfG(z,z') = \bfsfG_{\rm h}(z-z') - \bfsfG_{\rm h}(z +  z') + 2\left(\frac{ k_{\parallel}c}{n\omega}\right)^{2}g_{\rm h}(z+z') \hat {\bf z} \hat {\bf z},
\end{equation}
where the $({\bf k}_{\parallel},\omega)$-dependence of the Green functions was again suppressed.
To understand energy transfer rates near a mirror, we need to determine the Green function in the real-space representation, which is related to the previous equation by the inverse Fourier transform
%
\begin{equation}
\bfsfG({\bf r},{\bf r'},\omega) = \frac{1}{(2\pi)^{2}}\int\mbox{d}^{2}{\bf k}_{\parallel} \bfsfG({\bf k}_{\parallel},z,z',\omega) e^{i {\bf k}_{\parallel}\cdot ({\bm \rho} - {\bm \rho}')},
\end{equation}
%
where ${\bm \rho} = ({\bf x,y})$ and ${\bm \rho}' = ({\bf x',y'})$ so that ${\bf r}= ({\bm \rho},z)$.
Knowing that this inverse Fourier transform when applied to $\bfsfG_{\rm h}({\bf k}_{\parallel},z-z',\omega)$ leads to the expression~(\ref{Ghom_realspace}) also helps to evaluate the transform of $\bfsfG_{\rm h}({\bf k}_{\parallel},z+z',\omega)$.
Similarly, the third term on the right-hand side of Eq.~(\ref{Gmirrorassumof3terms}) transform analogous to the
$\hat {\bf z} \hat {\bf z}$-component of the homogeneous-medium Green tensor.
We thus find the Green function for an ideal mirror within a homogeneous medium as the sum of three terms:
%
\begin{eqnarray}\label{eq:G_for_ideal_mirror}
\bfsfG({\bf r},{\bf r'},\omega) & =& \bfsfG_{\rm h}({\bf r},{\bf r'},\omega) - \bfsfG_{\rm h}({\bm \rho},z+z',{\bm \rho}',0,\omega) \nonumber \\
&& + 2  G_{0}^{zz}({\bm \rho},z+z',{\bm \rho}',0,\omega)\hat {\bf z} \hat {\bf z}.
\end{eqnarray}
%
For the parallel configuration, we find
\begin{eqnarray}\label{muGmu_parallel}
{\bm \mu}^{\parallel}\cdot \bfsfG({\bf r}_{\rm a},{\bf r}_{\rm d},\omega)\cdot {\bm \mu}^{\parallel} & = & - \mu^{2} \frac{e^{ i n\omega r_{\rm da}/c}}{4 \pi r_{\rm da}} P(i n\omega r_{\rm da}/c) \nonumber \\
 & + & \mu^{2} \frac{e^{ i n\omega u/c}}{4 \pi u} P(i n\omega u/c),
\end{eqnarray}
where $r_{\rm da}$ is the donor-acceptor distance, $z$ the distance of both dipoles to the mirror, and $u \equiv [r_{\rm da}^{2} + (2 z)^2]^{1/2}$, and ${\bm \mu}^{\parallel} = \mu \hat y$ as in Fig.~\ref{Fig:situation}.
Inserting this result into Eq.~(\ref{wamplitude}) immediately gives the squared transfer amplitude $w({\bf r}_{\rm a},{\bf r}_{\rm d},\omega)$ of Eq.~(\ref{wamplitude}) for the parallel configuration.

For the perpendicular configuration we find
\begin{eqnarray}\label{muGmu_perpendicular}
& & {\bm \mu}^{\perp}\cdot \bfsfG({\bf r}_{\rm a},{\bf r}_{\rm d},\omega)\cdot {\bm \mu}^{\perp}  =  - \mu^{2} \frac{e^{ i n\omega r_{\rm da}/c}}{4 \pi r_{\rm da}} P(i n\omega r_{\rm da}/c) \\
 & & -  \mu^{2} \frac{e^{ i n\omega u/c}}{4 \pi u} \left[ P(i n\omega u/c) + 4\left(\frac{z}{u}\right)^{2} Q(i n\omega u /c)\right], \nonumber
\end{eqnarray}
with ${\bm \mu}^{\perp} = \mu \hat z$ as in Fig.~\ref{Fig:situation} and $u$ as in Eq.~(\ref{muGmu_parallel}), whereby the squared transfer amplitude of Eq.~(\ref{wamplitude}) is also determined for the perpendicular configuration.

For completeness, we also give single-emitter spontaneous-emission rates near the mirror (neglecting local-field effects~\cite{Schuurmans1998PRL} here and in the following).
For a dipole at a distance $z$ and oriented parallel to the mirror, we find from Eqs.~(\ref{gammadby2}) and ~(\ref{eq:G_for_ideal_mirror})
\begin{subequations}
\begin{equation}
\gamma_{\rm se}^{\parallel}(z,\omega) = \gamma_{\rm se, h}(\omega)\left\{ 1 - \frac{3}{2}\left[ \frac{\sin(\alpha)}{\alpha} + \frac{\cos(\alpha)}{\alpha^2} - \frac{\sin(\alpha)}{\alpha^3}\right]\right\},
\end{equation}
in terms of $\alpha = 2 n \omega z/c$ and the homogeneous-medium spontaneous-emission rate $\gamma_{\rm se, h} = \mu^{2} n \omega^{3}/(3 \pi \hbar \varepsilon_{0} c^{3})$, \emph{i.e.}, $n$ times the  spontaneous-emission rate in free space.
For a dipole emitter oriented normal to the mirror, the position-dependent spontaneous emission rate becomes
\begin{equation}
\gamma_{\rm se}^{\perp}(z,\omega) = \gamma_{\rm se, h}(\omega)\left\{1 - 3\left[ \frac{\cos\alpha}{\alpha^2} - \frac{\sin\alpha}{\alpha^{3}}\right]\right\}.
\end{equation}
\end{subequations}
In the limit $z \to 0$, the rate $\gamma_{\rm se}^{\parallel}(z,\omega)$ vanishes, while
$\gamma_{\rm se}^{\perp}(z,\omega)$ tends to $2 \gamma_{\rm se, h}$.
In the limit $z \to \infty$, both $\gamma_{\rm se}^{\parallel}(z,\omega)$ and $\gamma_{\rm se}^{\perp}(z,\omega)$ tend to the homogeneous-medium rate $\gamma_{\rm se, h}(\omega)$.
These two expressions are well known~\cite{Matloob:2000a}.
Here we see how these exact results also follow from our multiple-scattering approach; we will compare their spatial dependence with that of the analytically determined energy transfer rates obtained by the same approach.

\begin{figure}[tbp]
\includegraphics[width=1.0\columnwidth]{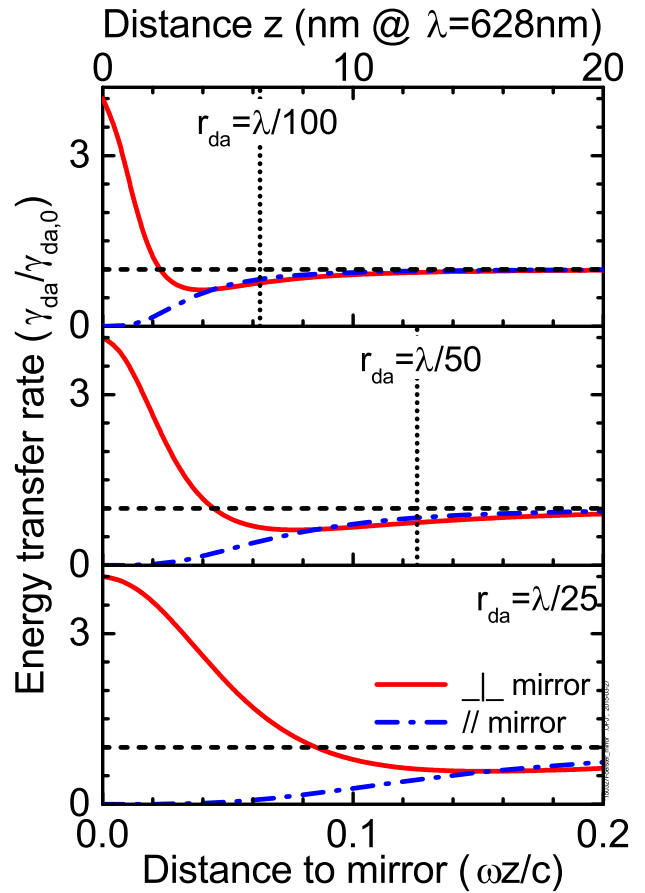}
\caption{(Color online)
Total energy transfer rate between a donor and an acceptor dipole, scaled to the free-space transfer rate, versus distance to the mirror, for the parallel and perpendicular configurations.
The lower abscissa gives the distance in scaled units, and the upper abscissa absolute distances at a wavelength $\lambda = (2\pi \cdot 100) {\rm nm} = 628 {\rm nm}$.
From top to bottom the three panels correspond to donor-acceptor spacings $r_{\rm da} = \lambda/100$, $\lambda/50$, $\lambda/25$, where dipole-mirror distances equal to $r_{\rm da}$ are marked by vertical dotted lines (off scale in the lowest panel).
}
\label{Fig:FRET_dist}
\end{figure}
%

\section{Energy transfer near a mirror: phenomenology}
Figure~\ref{Fig:FRET_dist} shows the total energy transfer rate between a donor and an acceptor as a function of distance $z$ to the mirror.
In this figure and all others below we use index $n=1$.
The panels show results for several donor-acceptor spacings $r_{\rm da} = \lambda/100$, $\lambda/50$, $\lambda/25$.
In all cases, the total energy transfer reveals a considerable dependence at short range.
In the limit of vanishing dipole-mirror distance ($z \rightarrow 0$), dipoles perpendicular to the mirror have a four-fold enhanced transfer rate compared to free space.
The factor four can be understood from the well-known method of image charges in electrodynamics: at a vanishing distance, each image dipole enhance the field two-fold, and since energy transfer invokes two dipoles, the total result is a four-fold enhancement.

With increasing dipole-mirror distance, the rate shows a minimum at a characteristic distance that is remarkably close to the donor-acceptor spacing $z \simeq r_{\rm da}$.
At larger distances $z > r_{\rm da}$, the transfer rate converges to the rate in the homogeneous medium.
In Appendix~\ref{App:Scaling} it is shown that this holds more generally: away from surfaces or other inhomogeneities, the FRET rate in an inhomogeneous medium scales increasingly as $\propto 1/({n}^{4}r_{\rm da}^{6})$, with $n$ the refractive index surrounding the donor-acceptor pair.
Figure~\ref{Fig:FRET_dist} also shows that in the limit of vanishing dipole-mirror distance ($z \rightarrow 0$), dipoles parallel to the mirror have an inhibited transfer rate.
This result can also be understood from the method of image charges, since each image dipole provides complete destructive interference in the limit of zero distance to the mirror.

With increasing dipole-mirror distance, the rate increases monotonously, and reaches half the free-space rate at a characteristic distance that is also remarkably close to the donor-acceptor spacing $z \simeq r_{\rm da}$.
At larger distances $z > r_{\rm da}$, the transfer rate tends to the homogeneous-medium rate.
It is remarkable that even in a simple system studied here, a dramatic modification of the energy transfer rate is feasible.
In other words, we can already conclude that the energy transfer rate between donor and acceptor is controlled by the distance to the mirror.
The open question is whether this control is mediated by the LDOS.

\begin{figure}[t]
\includegraphics[width=1.0\columnwidth]{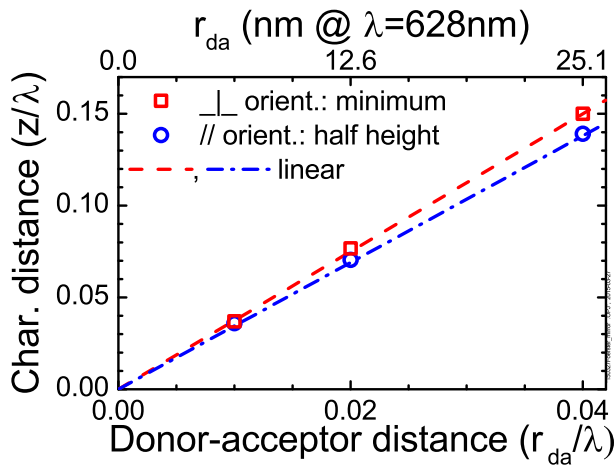}
\caption{(Color online)
Characteristic mirror separations $z/\lambda$ at which the total energy-transfer rate converges to the one in free-space.
For dipoles perpendicular to the mirror, the characteristic distance is shown at which the total energy transfer rate $\gamma_{DA}/\gamma_{DA, 0}$ has a minimum (see Fig.~\ref{Fig:FRET_dist}).
For dipoles parallel to the mirror, the characteristic distance is shown where $\gamma_{\rm da}/\gamma_{da, 0}$ equals $1/2$ (see Fig.~\ref{Fig:FRET_dist}).
The lines are linear fits through the origin with slopes $3.75$ and $3.45$, respectively.
}
\label{Fig:Char_Distances}
\end{figure}
%

Figure~\ref{Fig:Char_Distances} shows typical distances that characterize the distance dependent energy transfer rates in Figure~\ref{Fig:FRET_dist} versus donor-acceptor distance.
For the perpendicular configuration we plot the distance where the transfer rate has a minimum, and for the parallel case we plot the distance where the transfer rate equals $1/2$ of the free space rate.
Both characteristic distances increase linearly with the donor-acceptor distance with near-unity slope.
This behavior confirms that the distance dependence of the energy transfer rates in Figure~\ref{Fig:FRET_dist} occurs on length scales comparable to the donor-acceptor distance.

\begin{figure}[t]
\includegraphics[width=1.0\columnwidth]{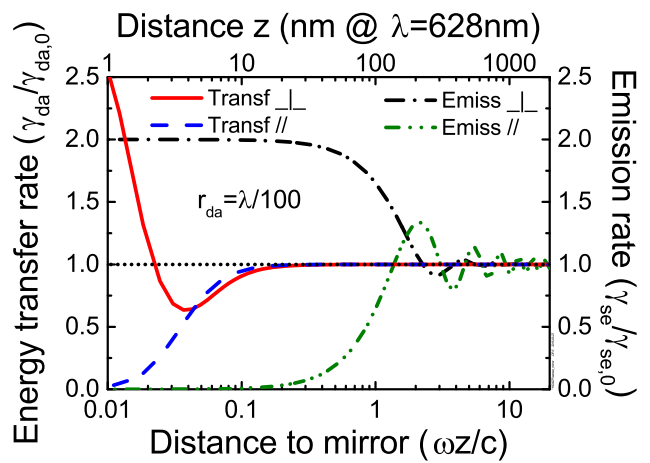}
\caption{(Color online) Comparison of donor-acceptor energy transfer rates $\gamma_{\rm da}$ and donor-only spontaneous emission rates $\gamma_{\rm se}$, as a function of the distance $z$ to the mirror.
The lower abscissa is the scaled distance, the top abscissa is the absolute distance for $\lambda = 628$ nm, both on a log scale.
The energy transfer is scaled by the free-space energy transfer rate $\gamma_{\rm da,0}$, the spontaneous emission by the free-space rate $\gamma_{\rm se,0}$.
Data are shown both for the parallel and for the perpendicular configurations.
For vanishing distance, $\gamma_{\rm da}/\gamma_{\rm da,0}$ is inhibited to 0 for the parallel and enhanced to 4 for the perpendicular configuration.
}
\label{Fig:100grad_FRET_dist}
\end{figure}
%

Figure~\ref{Fig:100grad_FRET_dist} shows the distance-dependence of the energy transfer rate in comparison to the spontaneous-emission rate.
The latter varies with distance to the mirror on length scales comparable to the wavelength of light, as first discovered by Drexhage~\cite{Drexhage1970JL}.
In contrast, the energy transfer rates vary on dramatically shorter length scales, about one-and-a-half (parallel configuration) to two (perpendicular configuration) \emph{orders of magnitude} smaller than the wavelength scale.

\begin{figure}[t]
\includegraphics[width=1.0\columnwidth]{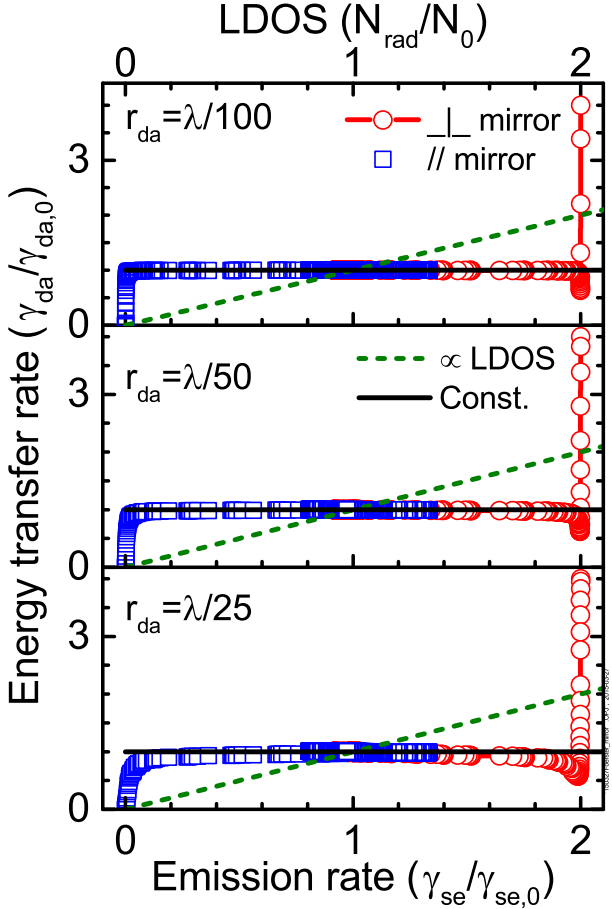}
\caption{(Color online) Parametric plots of scaled energy transfer rate versus scaled spontaneous-emission rate, for three donor-acceptor distances $r_{\rm da} = \lambda/100$, $\lambda/50$, $\lambda/25$ from top to bottom, respectively.
Data are obtained from Fig.~\ref{Fig:100grad_FRET_dist}.
Black horizontal lines are constant at $\gamma_{\rm da}/\gamma_{da,0} = 1$.
The green dashed lines are linear relations $\gamma_{\rm da} = \gamma_{\rm se}$.
}
\label{Fig:FRET_vs_G}
\end{figure}
%

To graphically investigate a possible relation between energy transfer rate and LDOS, Figure~\ref{Fig:FRET_vs_G} shows a parametric plot of the energy transfer rate as a function of (donor-only) spontaneous-emission rate, where each data point pertains to a certain distance $z$.
The top abscissa is the relative LDOS at the donor emission frequency that equals the relative emission rate.
The results at lower emission rate correspond mostly to the parallel dipole configurations in Figs.~\ref{Fig:FRET_dist} and~\ref{Fig:100grad_FRET_dist}, whereas the results at higher emission rate correspond to mostly to the perpendicular configurations in these figures.
For three donor-acceptor distances ($r_{\rm da} = \lambda/100$, $\lambda/50$, $\lambda/25$) Figure~\ref{Fig:FRET_vs_G} shows that the energy transfer rate is independent of the emission rate and the LDOS over nearly the whole range, in agreement with conclusions of Refs.~\cite{Dood2005PRB, Blum2012PRL, Rabouw2014NC, Konrad2015NS}.
The energy transfer decreases fast near the low emission rate edge and increases fast near the high emission rate edge, both of which correspond to distances very close to the mirror (\emph{cf.} Fig.~\ref{Fig:FRET_dist}).
From Figure~\ref{Fig:FRET_vs_G} it is readily apparent that the energy transfer rate does not increase linearly with the LDOS, leave alone quadratically, as proposed previously.
The absence of a correlation between energy transfer rate and LDOS that is phenomenologically shown here is one of our main results, and will be theoretically discussed in the remainder of this paper.

\section{Energy transfer versus LDOS}\label{Sec:Foerster_versus_total}
As is well-known, not all energy transfer is F{\" o}rster energy transfer.
To find what fraction of the energy transfer corresponds to F{\" o}rster energy transfer, we express the Green function in terms of the complete set of optical eigenmodes ${\bf f}_{\lambda}$ that satisfy the wave equation
%
\begin{equation}\label{eqformodes}
 -\nabla\times\nabla \times {\bf f}_{\lambda}({\bf r}) + \varepsilon({\bf r})(\omega_{\lambda}/c)^{2}{\bf f}_{\lambda}({\bf r}) = 0,
\end{equation}
%
with positive eigenfrequencies $\omega_{\lambda} > 0$.
The Green function, being the solution of Eq.~(\ref{eqforG}), can be expanded in terms of these mode functions ${\bf f}_{\lambda}$.
An important property of this expansion can now be obtained from Ref.~\onlinecite{Wubs:2004a} (in particular by combining Eqs.~(21) and (22) of~\cite{Wubs:2004a}), namely that the Green function can be written as the following sum of three terms:
%
\begin{eqnarray}\label{Gin 3terms}
{\bfsfG}({\bf r},{\bf r'},\omega)  & = &
\underbrace{
c^{2} \sum_{\lambda} \frac{{\bf f}_{\lambda}({\bf r}) {\bf f}_{\lambda}^{*}({\bf r'})}{(\omega + \mathrm{i}\eta)^{2}-\omega_{\lambda}^{2}}
}
 \\
& & \qquad\qquad  {\bfsfG}_{\rm R}  \nonumber \\
&&\underbrace{
-\left(\frac{c}{\omega}\right)^{2}\sum_{\lambda} {\bf f}_{\lambda}({\bf r}) {\bf f}_{\lambda}^{*}({\bf r'})
}
+
\frac{(c/\omega)^2}{\varepsilon({\bf r})}\delta({\bf r}-{\bf r'}){\bfsfI}. \nonumber \\
& & \qquad\qquad {\bfsfG}_{\rm S}  \nonumber
\end{eqnarray}
%
Since the Green function controls the energy transfer rate (see Eq.~(\ref{wamplitude})), it is relevant to discern energy transfer processes corresponding to these terms.
The first term in Eq.~(\ref{Gin 3terms}) denoted ${\bfsfG}_{\rm R}$ corresponds to resonant dipole-dipole interaction (RDDI), the radiative process by which the donor at position ${\bf r}$ emits
a field that is then received by the acceptor at position ${\bf r'}$.
The name `resonant' describes that photon energies close to the donor and acceptor resonance energy are the most probable energy transporters, in line with the denominator $(\omega + \mathrm{i}\eta)^{2}-\omega_{\lambda}^{2}$ of this first term.
The second term in~(\ref{Gin 3terms}) called ${\bfsfG}_{\rm S}$ corresponds to the static dipole-dipole interaction (SDDI) that also causes energy transfer from donor to acceptor, yet by virtual intermediate processes (see also Sec. V.B of Ref.~\onlinecite{Wubs:2004a}).
As explained below, it is this SDDI that gives rise to the FRET rate that characteristically scales as $r_{\rm da}^{-6}$ in homogeneous media and dominates the total energy transfer for strongly subwavelength donor-acceptor separations.
The third term in Eq.~(\ref{wamplitude}) is proportional to the Dirac delta function $\delta({\bf r} - {\bf r'})$.
Since ${\bf r}\ne {\bf r'}$ in case of energy transfer, this contribution vanishes.

The fact that the Green function can be written as the sum of three terms as in Eq.~(\ref{Gin 3terms}) is important, and implies that for arbitrary environments the static part of the Green function can be obtained from the total Green function by the following limiting procedure (for ${\bf r}\ne {\bf r'}$)
%
\begin{equation}\label{GS_as_limit}
\bfsfG_{\rm S}({\bf r}, {\bf r'},\omega) = \frac{1}{\omega^{2}}\lim_{\omega \rightarrow 0} \omega^{2} \bfsfG({\bf r}, {\bf r'},\omega),
\end{equation}
%
which provides a justification of our use of the term `static'.
As an important test, selecting in this way the static part of the Green function of a homogeneous medium~(\ref{Ghom_realspace}) indeed gives that only
%
\begin{equation}\label{eq:Gshom}
\bfsfG_{\rm h, S}({\bf r}_{1}, {\bf r}_{2},\omega) = \frac{c_{0}^{2}}{4 \pi n^{2} \omega^{2} r^{3}}\left( \bfsfI - 3 \hat {\bf r} \hat {\bf r}\right),
\end{equation}
%
with ${\bf r} = {\bf r}_{1} - {\bf r}_{2}$ contributes to F{\" o}rster energy transfer, and not the terms of $\bfsfG_{\rm h}$ that vary as $1/r$ and $1/r^{2}$.
Incidentally and by contrast, for general inhomogeneous media the static Green function does not only depend on the distance between donor and emitter, but rather on the absolute positions of both donor and acceptor in the medium.

Having thus defined F{\" o}rster energy transfer as that part of the total energy transfer that is mediated by the static dipole-dipole interaction, we can now also define the square of the F{\"o}rster transfer amplitude, in analogy to Eq.~(\ref{wamplitude}),  by
%
\begin{equation}\label{wamplitudefoerster}
w_{\rm F}({\bf r}_{\rm a},{\bf r}_{\rm d}, \omega) = \frac{2\pi}{\hbar^{2}} \left(\frac{\omega^{2}}{\varepsilon_{0} c^{2}}\right)^{2} \; |{\bm  \mu}_{\rm a}^{*}\cdot {\bfsfG}_{\rm S}({\bf r}_{\rm a},{\bf r}_{\rm d},\omega)\cdot {\bm  \mu}_{\rm d}|^2.
\end{equation}
%
This appears similar to Eq.~(\ref{wamplitude}), yet with the total Green function $\bfsfG$ replaced by its static part $\bfsfG_{\rm S}$, as defined in Eq.~(\ref{Gin 3terms}) and computed in Eq.~(\ref{GS_as_limit}).
The FRET rate $\gamma_{\rm F}$ is then obtained by substituting $w_{\rm F}({\bf r}_{\rm a},{\bf r}_{\rm d}, \omega)$ for $w({\bf r}_{\rm a},{\bf r}_{\rm d}, \omega)$ into Eq.~(\ref{energy-transfer-rate}), giving:
\begin{equation}\label{eq:FRETrate}
\gamma_{\rm F}({\bf r}_{\rm a},{\bf r}_{\rm d})  =  \int_{-\infty}^{\infty} \mbox{d}\omega\,\sigma_{\rm a}(\omega)\, w_{\rm F}({\bf r}_{\rm a},{\bf r}_{\rm d}, \omega)\, \sigma_{\rm d}(\omega).
\end{equation}
Here we arrive at an important simplification in the description of F{\"o}rster transfer in inhomogeneous media, by noting that from Eqs.~(\ref{GS_as_limit}) and ({\ref{wamplitudefoerster}), the quantity $w_{\rm F}({\bf r}_{\rm a},{\bf r}_{\rm d}, \omega)$ is {\em independent of frequency} $\omega$.
The FRET rate $\gamma_{\rm F}$ is then given by the simple relation
%
\begin{equation}\label{eq:FRETrate2}
\gamma_{\rm F}({\bf r}_{\rm a},{\bf r}_{\rm d})  = w_{\rm F}({\bf r}_{\rm a},{\bf r}_{\rm d})\,\int_{-\infty}^{\infty} \mbox{d}\omega\,\sigma_{\rm a}(\omega)\,  \sigma_{\rm d}(\omega).
\end{equation}
%
While this expression looks similar to the approximate expression for the total energy transfer rate (Eq.~\ref{energy-transfer-rate-approx}), we emphasize as a first point that Eq.~(\ref{eq:FRETrate2}) is an {\em exact} expression for the FRET rate, even for broad donor and acceptor spectra.
A second crucial point is that the spectral overlap integral in Eq.~(\ref{eq:FRETrate2}) is the same for {\em any} nanophotonic environment \footnote{Let us recall here that $\sigma_{\rm a}(\omega)$ and $\sigma_{\rm d}(\omega)$ are the donor's emission spectrum and acceptor's absorption spectrum in {\em free space}, see Eq.~(\ref{energy-transfer-rate}) and Refs.~\cite{May:2000a, Dung2002PRA}).}.
All effects of the nondispersive inhomogeneous environment are therefore contained in the frequency-independent prefactor $w_{\rm F}({\bf r}_{\rm a},{\bf r}_{\rm d})$.
In other words, while there is an effect of the nanophotonic environment on the FRET rate  (see Fig.~\ref{Fig:FRET_dist}), this effect depends only on the donor and acceptor positions but does not depend on the resonance frequencies of the donor and acceptor (for constant medium-independent overlap integral in Eq.~(\ref{eq:FRETrate2})).
If the FRET rate does not depend on the donor and acceptor frequencies, then the FRET rate can not be a function of the LDOS at these particular frequencies.
A third crucial point is that this conclusion is valid for {\em any} photonic environment that is lossless and weakly dispersive in the frequency range where the donor and acceptor spectra overlap, hence this conclusion is not limited to an ideal mirror.

For homogeneous media it is well known that F{\"o}rster energy transfer dominates the total energy transfer at strongly sub-wavelength distances, and we will now see that this is also the case in inhomogeneous media, again taking the ideal mirror as an example.
The total energy transfer near an ideal mirror depends on the total Green function as given in Eqs.~(\ref{muGmu_parallel}) and~(\ref{muGmu_perpendicular}) for the two dipole configurations (\emph{cf.} Fig.~\ref{Fig:situation}).
For the donor and acceptor near the mirror in the parallel configuration, we use the procedure of Eq.~(\ref{GS_as_limit}) and obtain for the static parts
\begin{equation}\label{mu_Gs_mu_parallel}
{\bm \mu}^{\parallel}\cdot\bfsfG_{\rm S}({\bf r}_{\rm a}, {\bf r}_{\rm d},\omega)\cdot {\bm \mu}^{\parallel} = \frac{\mu^{2} c^{2}}{4 \pi n^{2}\omega^{2}}\left\{ \frac{1}{r_{\rm da}^{3}} - \frac{1}{(\sqrt{r_{\rm da}^2 + 4 z^{2}})^3}\right\},
\end{equation}
while for the perpendicular configuration we find
\begin{eqnarray}\label{mu_Gs_mu_perpendicular}
&& {\bm \mu}^{\perp}\cdot\bfsfG_{\rm S}({\bf r}_{\rm a}, {\bf r}_{\rm d},\omega)\cdot {\bm \mu}^{\perp} = \frac{\mu^{2} c^{2}}{4 \pi n^{2}\omega^{2}}\left\{ \frac{1}{r_{\rm da}^{3}} + \nonumber \right. \\ && \left. \frac{1}{(\sqrt{r_{\rm da}^2 + 4 z^{2}})^3}\left( 1 - 3\frac{4 z^{2}}{r_{\rm da}^{2} + 4 z^{2}}\right)\right\}.
\end{eqnarray}
We note that in both cases the static interaction in a homogeneous medium is recovered for FRET pairs at distances to the mirror much larger than the donor-acceptor distance $z \gg r_{\rm da}$.
This agrees with the large-distance limits shown in Figs.~\ref{Fig:FRET_dist} and~\ref{Fig:100grad_FRET_dist}.
The spatial dependence of the F{\"o}rster transfer amplitude of Eq.~(\ref{wamplitudefoerster}) and of the FRET rate in Eq.~(\ref{eq:FRETrate2}) is hereby determined for both configurations.

\begin{figure}[t]
\includegraphics[width=1.0\columnwidth]{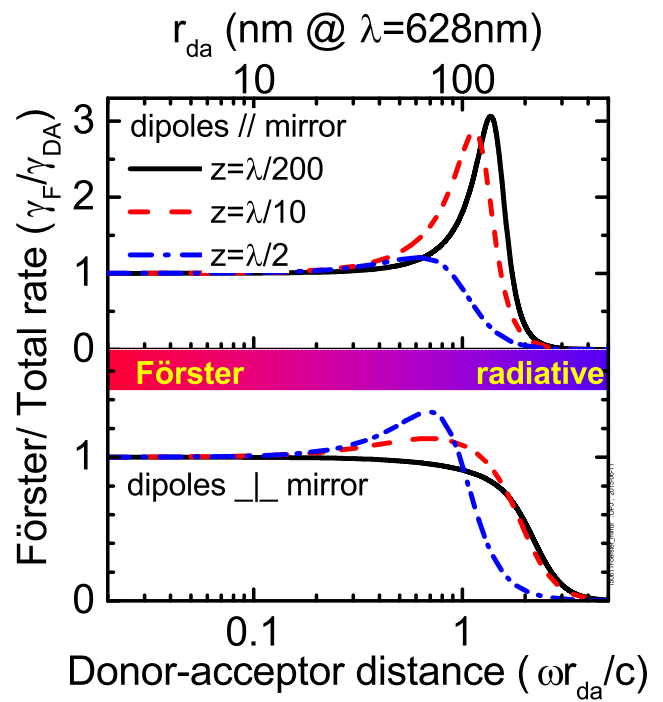}
\caption{(Color online)
F{\"o}rster resonance energy transfer rate scaled to the total energy transfer rate $(\gamma_{\rm F}/\gamma_{\rm da})$ versus donor-acceptor distance $r_{\rm da}$ for three distances $z$ of donor and acceptor to the mirror.
The upper panel is for dipoles parallel to the mirror, the lower panel for dipoles perpendicular to the mirror.
Note the logarithmic $r_{\rm da}$, with dimensionless scaled values on the lower abscissa and absolute distance in nanometers on the upper abscissa for $\lambda = 628{\rm nm}$.
}
\label{Fig:interpret_vs_DA-distance}
\end{figure}
%
In Figure~\ref{Fig:interpret_vs_DA-distance} we display the ratio of the FRET rate and the total energy transfer rate as a function of donor-acceptor distance, for three distances $z$ of the FRET pair to the mirror, and for both dipole configurations, taking $n=1$.
For the total rate we use the narrow bandwidth assumption of Eq.~(\ref{energy-transfer-rate-approx}).
Irrespective of the distance to the mirror and of dipole configuration, the total energy transfer rate equals the FRET rate for all practical purposes, as long as $(r_{\rm da} \omega/c \ll 1)$, in other words for strongly sub-wavelength donor-acceptor distances.
Intriguingly, with increasing donor-acceptor distance beyond typical FRET distances, the ratio of the two rates exceeds unity.
Since the total rate $\gamma_{\rm da}$ equals the absolute square of the sum of static and resonant transfer amplitudes, and the two amplitudes interfere destructively in this intermediate distance range, the FRET (static) rate can indeed dominate the total energy transfer rate.
At large donor-acceptor distances $(r_{\rm da} \omega/c \gg 1)$, the FRET rate decreases much faster with distance than the total transfer rate, similar as in homogeneous media.
In this large-distance range, the energy transfer is radiative: the donor emits a photon that is absorbed by the acceptor.

%
\begin{figure}[t]
\includegraphics[width=1.0\columnwidth]{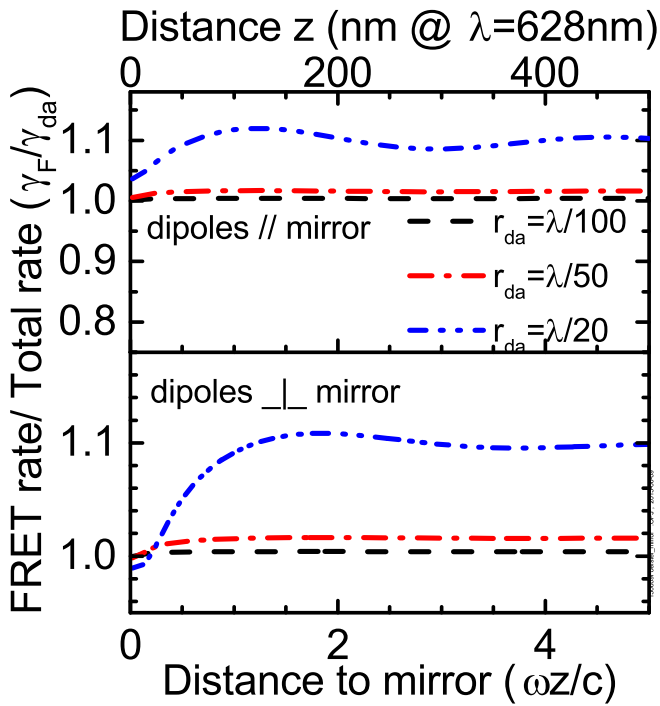}
\caption{(Color online)
FRET rate $\gamma_{\rm F}$ divided by the total energy transfer rate $\gamma_{\rm da}$, versus distance to the mirror, for three values of the donor-acceptor distance $r_{\rm da}$.
The lower abscissa is the dimensionless reduced distance, the upper abscissa is the absolute distance in nanometer for $\lambda = 628 {\rm nm}$.
Upper panel: parallel dipole configuration; lower panel: perpendicular dipole configuration.
}
\label{Fig:interpret_vs_mirror-distance}
\end{figure}
%
Figure~\ref{Fig:interpret_vs_mirror-distance} is complementary to the previous one in the sense that here the FRET rate is plotted versus distance to the mirror $z$ for several donor-acceptor distances $r_{\rm da}$, and for both dipole configurations.
We again show the ratio of the FRET rate and the total transfer rate (using Eq.~(\ref{energy-transfer-rate-approx}) for $\gamma_{\rm da}$).
At donor-acceptor distances $r_{\rm da} = \lambda/100$ and $r_{\rm da} = \lambda/50$, typical for experimental situations, we clearly see that FRET dominates the total energy transfer rate, independent of the distance to the mirror.
At least $98 \%$ of the total energy transfer rate consists of the FRET rate.
Even for a large donor-acceptor distance $r_{\rm da} = \lambda /20$ that is much larger than in most experimental cases (that is, $r_{\rm da} = 31 {\rm nm}$ at $\lambda = 628~{\rm nm}$), the FRET rate and the total rate differ by only some ten percent.
Thus, Figures~\ref{Fig:interpret_vs_DA-distance} and~\ref{Fig:interpret_vs_mirror-distance} illustrate that in the nanophotonic case near an ideal mirror, the FRET greatly dominates the total energy transfer at strongly sub-wavelength donor-acceptor distances, similar as in the well-known case of homogeneous media.
Therefore, we conclude that not only is there no correlation between the LDOS and the total transfer rate, there is also no correlation between the LDOS and the FRET rate either.

\section{Energy transfer in terms of a frequency-integrated LDOS}\label{Sec:FLDOS}
Our general derivation of the FRET rate in a weakly dispersive nanophotonic medium~(Eq.~\ref{eq:FRETrate2}) has convincingly shown that the FRET rate has no dependence on the local density of optical states evaluated at the donor's resonance frequency.
In this section we will insist on establishing a link between the FRET rate and the LDOS, if only to counter the argument that we are from the outset biased against such a relation. Interestingly, the relation that we derive will also shed a new light on efforts to control the FRET rate by LDOS engineering.

We start with the mode expansion of the Green function in Eq.~(\ref{Gin 3terms}) to derive a useful new expression, relating the F{\"o}rster transfer rate to a frequency-integral over $\mbox{Im}[{\bfsfG}]$.
We use the fact that ${\bfsfG}_{\rm S}({\bf r},{\bf r'},\omega)$ is real-valued, as is proven in Ref.~\cite{Wubs:2003a}.
Thus the imaginary part of the Green function is equal to $\mbox{Im}[{\bfsfG}_{\rm R}]$ and the mode expansion of $\mbox{Im}[{\bfsfG}]$ becomes
%
\begin{equation}\label{ImGexpansion}
\mbox{Im}[{\bfsfG}({\bf r},{\bf r'},\omega)] = - \frac{\pi c^{2}}{2 \omega}\sum_{\lambda} {\bf f}_{\lambda}({\bf r}) {\bf f}_{\lambda}^{*}({\bf r'}) \delta(\omega - \omega_{\lambda}),
\end{equation}
%
with $\omega > 0$.
We note that only modes with frequencies $\omega_{\lambda} = \omega$ show up in this mode expansion of $\mbox{Im}[{\bfsfG}]$.
This can also be seen in another way: the defining equation for the Green function Eq.~(\ref{eqforG}) implies that the {\em imaginary} part of the Green function satisfies the same source-free equation~(\ref{eqformodes}) as the subset of modes ${\bf f}_{\lambda}({\bf r})$ for which the eigenfrequency $\omega_{\lambda}$ equals $\omega$.
Therefore, $\mbox{Im}[{\bfsfG}({\bf r},{\bf r}',\omega)]$ can be completely expanded in terms of only those degenerate eigenmodes.
The mode expansion~(\ref{ImGexpansion}) is indeed a solution of Eq.~(\ref{eqformodes}).
If we multiply the right-hand side of Eq.~(\ref{ImGexpansion}) with $\omega$ and then integrate over $\omega$, we obtain as one of our major results an exact identity for the static Green function ${\bfsfG}_{\rm S}$
%
\begin{equation}\label{newformula}
{\bfsfG}_{\rm S}({\bf r}_{\rm a},{\bf r}_{\rm d},\omega) = \frac{2}{\pi \omega^{2}}\int_{0}^{\infty}\mbox{d}\omega_{1}\,\omega_{1}\,\mbox{Im}[{\bfsfG}({\bf r}_{\rm a},{\bf r}_{\rm d},\omega_{1})].
\end{equation}
%
This identity is valid for a general nanophotonic medium in which material dispersion can be neglected.
Eq.~(\ref{newformula}) was derived using a complete set of modes, yet does not depend on the specific set of modes used.
When inserting this identity into Eq.~(\ref{wamplitudefoerster}), we obtain the F{\"o}rster transfer rate $w_{\rm S}(\omega)$ and hence the transfer rate $\gamma_{\rm F}$ of Eq.~(\ref{eq:FRETrate}) in terms of the imaginary part of the Green function.
While this is somewhat analogous to the well-known expression for the spontaneous-emission rate~(Eq.~\ref{gammadby2}), there are two important differences:
The first difference between Eq.~(\ref{newformula}) for F{\"o}rster energy transfer and Eq.~(\ref{gammadby2}) for spontaneous emission in terms of $\mbox{Im}[{\bfsfG}]$ is of course that Eq.~(\ref{newformula}) is an {\em integral over all positive frequencies}.
The second main difference is that in Eq.~(\ref{newformula}) the Green function $\mbox{Im}[{\bfsfG}({\bf r}_{\rm a},{\bf r}_{\rm d},\omega_{1})]$ appears with {\em two} position arguments - one for the donor and one for the acceptor - instead of only one position as in the spontaneous emission rate.
A major advantage of an expression in terms of $\mbox{Im}[{\bfsfG}]$ is that $\mbox{Im}[{\bfsfG}]$ does not diverge for ${\bf r}_{\rm a}\rightarrow {\bf r}_{\rm d}$, in contrast to $\mbox{Re}[{\bfsfG}]$.

In Appendix~\ref{App:two_tests} we verify and show explicitly that the identity in Eq.~(\ref{newformula}) holds both in homogeneous media as well as for the nanophotonic case of arbitrary positions near an ideal mirror.
The identity in Eq.~(\ref{newformula}) is important since it is our goal in this section to explore whether FRET rates are functionally related to the LDOS in any way.
With Eq.~(\ref{newformula}), both quantities can now be expressed in terms of the imaginary part of the Green function, which brings us considerably closer to the goal.

We now use Eq.~(\ref{newformula}) to derive an approximate expression $\bfsfG_{\rm S}^{\rm (L)}$ for the static Green function $\bfsfG_{\rm S}$ that allows us to relate the F{\"o}rster transfer rate to the frequency-integrated LDOS.
Our approximation is motivated by the fact that $\mbox{Im}[\bfsfG({\bf r}_{\rm d} - {\bf r}_{\rm a},\omega)]$ for homogeneous media (based on Eq.~(\ref{Ghom_realspace})) varies appreciably only for variations in the donor-acceptor distance $r_{\rm da}$ on the scale of the wavelength of light, typically $r_{\rm da} \simeq \lambda_{0} = 500\,{\rm nm}$ (with $\lambda_{0}=2\pi c/\omega_{0}$).
From Eq.~(\ref{eq:G_for_ideal_mirror}) it follows that the same holds true for $\mbox{Im}[\bfsfG({\bf r}_{\rm d},{\bf r}_{\rm a},\omega)]$ for the ideal mirror.
In contrast, F{\"o}rster energy transfer occurs on a length scale of $r_{\rm da} \simeq 5$ nm, typically a hundred times smaller.
Motivated by these considerations, we approximate $\mbox{Im}[{\bfsfG}({\bf r}_{\rm a},{\bf r}_{\rm d},\omega_{1})]$ in the integrand of Eq.~(\ref{newformula}) by the zeroth-order Taylor approximation $\mbox{Im}[{\bfsfG}({\bf r}_{\rm d},{\bf r}_{\rm d},\omega_{1})]$.
The accuracy of this approximation depends on the optical frequency $\omega$.
The approximation will not hold for all frequencies that are integrated over, and becomes worse for higher frequencies.
But it appears that we can make an accurate approximation throughout a huge optical bandwidth $0 \le \omega_{1} \le \Omega$.
If we choose $\Omega = 10\omega_{0}$, \emph{i.e.}, a frequency bandwidth all the way up to the vacuum ultraviolet (VUV), then $\mbox{Im}[{\bfsfG}({\bf r}_{\rm a},{\bf r}_{\rm d},\omega_{1})]$ will only deviate appreciably from $\mbox{Im}[{\bfsfG}({\bf r}_{\rm d},{\bf r}_{\rm d},\omega_{1})]$ for donor-acceptor distances $r_{\rm da} > \lambda_{0}/10$, which is in practice of the order of $50\,{\rm nm}$, much larger than typical donor-acceptor distances in F{\"o}rster transfer experiments.
We obtain the expression for the approximate static Green function $\bfsfG_{\rm S}^{\rm (L)}$ as
%
\begin{eqnarray}\label{newformula2}
{\bfsfG}_{\rm S}^{\rm (L)}({\bf r}_{\rm a},{\bf r}_{\rm d},\omega) &=& \frac{2}{\pi \omega^{2}}\int_{0}^{ \Omega}\mbox{d}\omega_{1}\,\omega_{1}\,\mbox{Im}[{\bfsfG}({\bf r}_{\rm d},{\bf r}_{\rm d},\omega_{1})]
\nonumber \\
&+&\frac{2}{\pi \omega^{2}}\int_{\Omega}^{\infty}\mbox{d}\omega_{1}\,\omega_{1}\,\mbox{Im}[{\bfsfG}({\bf r}_{\rm a},{\bf r}_{\rm d},\omega_{1})].
\end{eqnarray}
%
Here the first term is recognized to be an integral of the LDOS over a large frequency bandwidth, ranging from zero frequency (or `DC') to high frequencies in the VUV range.
While the specific value of $\Omega$ does not matter much, it is important that $\Omega$ can be chosen much greater than optical frequencies, while the inequality $n({\bf r}_{\rm d}) \Omega r_{\rm da}/c \ll 1$ still holds.
Within this approximation, the F{\"o}rster transfer rate can be related to the frequency-integrated LDOS: if we replace ${\bfsfG}_{\rm S}$ by ${\bfsfG}_{\rm S}^{\rm (L)}$ in Eq.~(\ref{wamplitudefoerster}) for $w_{\rm F}$, thereby obtaining the LDOS approximation $w_{\rm F}^{\rm (L)}({\bf r}_{\rm a},{\bf r}_{\rm d},\omega) =  |{\bm  \mu}_{\rm a}^{*}\cdot {\bfsfG}_{\rm S}^{\rm (L)}({\bf r}_{\rm a},{\bf r}_{\rm d},\omega)\cdot {\bm  \mu}_{\rm d}|^2$, and approximate  $w_{\rm F}$ in Eq.~(\ref{eq:FRETrate}) by this $w_{\rm F}^{\rm (L)}$, we obtain the relation between the approximate FRET rate $\gamma_{\rm F}^{\rm (L)}$ and the LDOS to be \footnote{In the symbol $\gamma_{\rm F}^{\rm (L)}$ the superscript L is meant to indicate a FRET rate in terms of the LDOS.}
%
\begin{figure}[t]
\includegraphics[width=1.0\columnwidth]{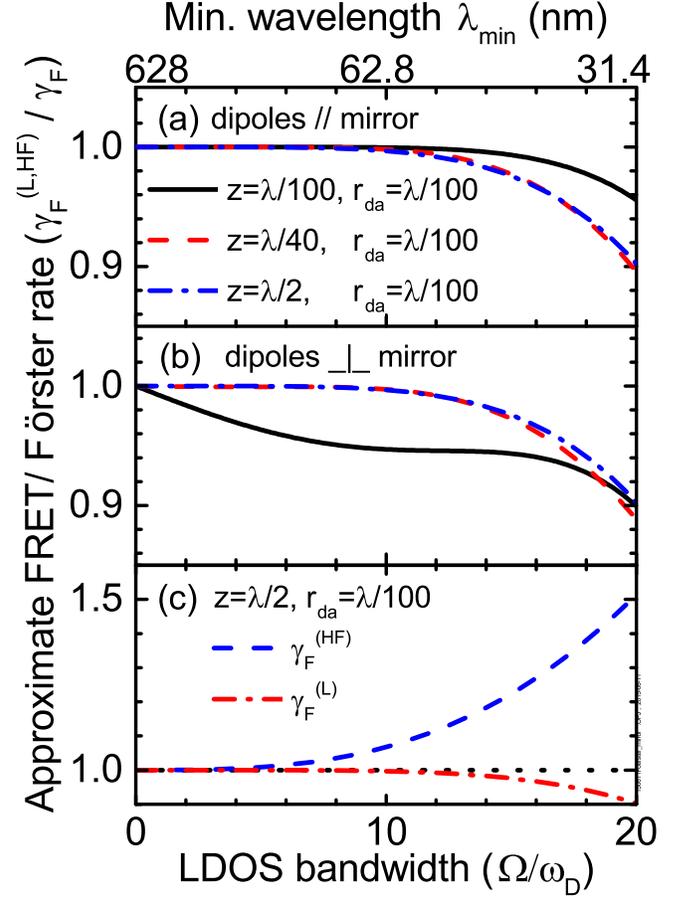}
\caption{(Color online)
LDOS-approximated FRET rate $\gamma_{\rm F}^{\rm (L)}$ (Eq.~(\ref{eq:FRETrateLDOS})) normalized to the exact FRET rate $\gamma_{\rm F}$ (Eq.~(\ref{eq:FRETrate})) versus the bandwidth $\Omega$ of the LDOS-frequency integral.
Lower abscissa: $\Omega$ scaled by the donor frequency $\omega_{\rm d}=2\pi c/\lambda$.
Upper abscissa: minimum wavelength $\lambda_{\rm min}=2\pi c/\Omega$ for $\lambda = 628{\rm  nm }$.
Black full curves are for dipole-to-mirror distance $z = \lambda/100$, red dashed curves for $z = \lambda/40$, and blue dashed-dotted curves for $z = \lambda/2$, all curves are for a donor-acceptor distance $r_{\rm da} = \lambda/100$.
(a) Parallel dipole configuration; (b) perpendicular dipole configuration.
(c) Comparison of the LDOS-approximation~(Eq.~(\ref{eq:FRETrateLDOS})) and the high-frequency approximation~(Eq.~(\ref{newformula_HF})) of the FRET rate as a function of LDOS bandwidth $\Omega$.
Rates are scaled to the exact FRET rate, and the distance to the mirror and the donor-acceptor distance are fixed.
}
\label{Fig:cut-off_freq}
\end{figure}
%
%
\begin{equation}\label{eq:FRETrateLDOS}
\gamma_{\rm F}^{\rm (L)} = \int_{-\infty}^{\infty} \mbox{d}\omega\,\sigma_{\rm a}(\omega)\, w_{\rm F}^{\rm (L)}(\omega)\, \sigma_{\rm d}(\omega).
\end{equation}
%
In Figure~\ref{Fig:cut-off_freq} we verify the accuracy of the LDOS-approximated FRET rate $\gamma_{\rm F}^{\rm (L)}$ near the ideal mirror, by varying the frequency bandwidth $\Omega$ over which we make the approximation.
The required frequency integrals of Eq.~(\ref{newformula2}) are calculated analytically in Appendix~\ref{App:accuracy_for_mirror}.
In Fig.~\ref{Fig:cut-off_freq} we see that for both dipole configurations, the approximate FRET rate indeed tends to the exact rate for vanishing $\Omega$.
For $\Omega$ up to $10 \omega_{\rm d}$, the approximate rate is very close to the exact one, to within $5 \%$.
Even at higher frequencies, up to $\Omega  = 20 \omega_{\rm d}$, the approximate FRET rate is within $10 \%$ of the exact rate, as anticipated on the basis of our general considerations above.

The validity of the approximate FRET rate $\gamma_{\rm F}^{\rm (L)}$ improves when the donor-acceptor distance $r_{\rm da}$ is reduced, since the spatial zero-order Taylor expansion of ${\rm Im}[\bfsfG]$ is then a better approximation.
We can also improve the approximation by reducing the frequency bandwidth $\Omega$ in which we make the Taylor approximation.
Both trends are indeed found in Appendix~\ref{App:accuracy_for_homogeneous_medium} where $\gamma_{\rm F}^{\rm (L)}$ is calculated for the homogeneous medium.
In the limit of a vanishing frequency bandwidth ($\Omega \rightarrow 0$), the approximate F{\"o}rster transfer rate $\gamma_{\rm F}^{\rm (L)}$ reduces to the exact F{\"o}rster transfer rate $\gamma_{\rm F}$ of Eq.~(\ref{eq:FRETrate}).

%
\begin{figure}[t]
\includegraphics[width=1.0\columnwidth]{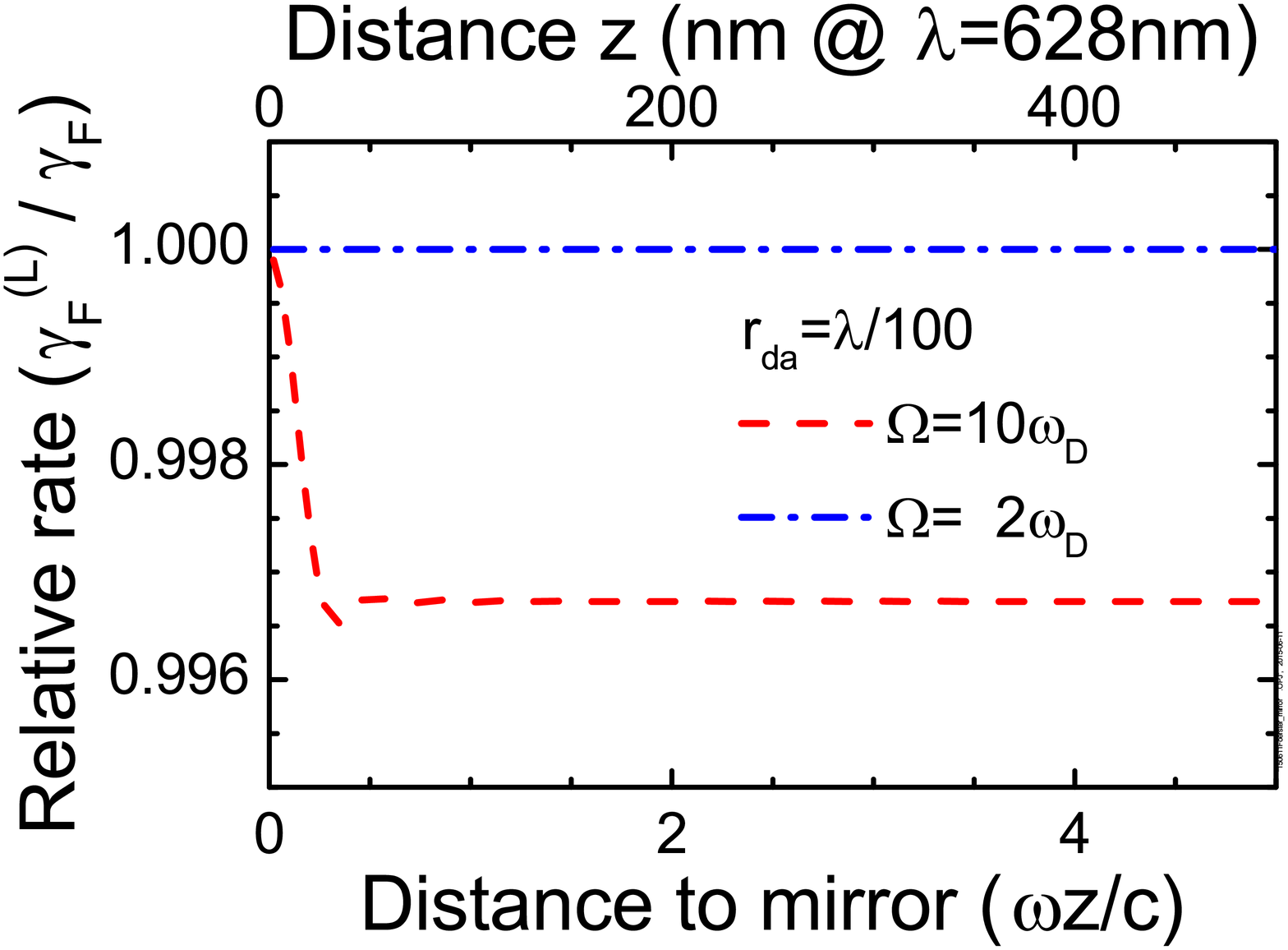}
\caption{(Color online)
LDOS-approximated FRET rate $\gamma_{\rm F}^{\rm (L)}$ (Eq.~(\ref{eq:FRETrateLDOS})) normalized to the exact FRET rate $\gamma_{\rm F}$ (Eq.~(\ref{eq:FRETrate})) versus (scaled) distance to the mirror, for an LDOS frequency bandwidth up to $\Omega = 10 \omega_{0}$ (red dashed curve), and up to $\Omega = 2 \omega_{0}$ (blue dashed-dotted curve).
The donor-acceptor distance is $r_{\rm da} = \lambda/100$.
}
\label{Fig:Taylor_vs_distance}
\end{figure}
%
To verify that the approximate FRET rates shown in Fig.~\ref{Fig:cut-off_freq} were not `lucky shots' for the chosen fixed distances to the mirror, we study in the complementary Figure~\ref{Fig:Taylor_vs_distance}(a) the accuracy of $\gamma_{\rm F}^{\rm (L)}$ as a function of distance to the mirror $z$, for a constant LDOS bandwidth $\Omega = 10 \omega_{\rm d}$.
The figure clearly shows the great accuracy of the LDOS approximation, irrespective of the distance $z$ of the FRET pair to the ideal mirror.
For a narrower bandwidth of $\Omega = 2 \omega_{\rm d}$, the accuracy is even better, as expected.

At this point, one may naively conclude from Figures~\ref{Fig:cut-off_freq} and~\ref{Fig:Taylor_vs_distance} that the FRET rate is intimately related to an integral over the LDOS.
This is too rash, however, because the corresponding approximate relation~Eq.~(\ref{eq:FRETrateLDOS}) consists of two integrals, where only one of them is an integral over the LDOS, while the other is a high-frequency integral of the complex part of the Green function.
Thus the relevant question becomes: what happens if we make a cruder approximation to the FRET rate by simply removing the LDOS integral?
Instead of Eq.~(\ref{newformula2}) we then use the high-frequency approximation (HF) to the static Green function ${\bfsfG}_{\rm S}^{\rm (HF)}$:
%
\begin{equation}\label{newformula_HF}
{\bfsfG}_{\rm S}^{\rm (HF)}({\bf r}_{\rm a},{\bf r}_{\rm d},\omega) = \frac{2}{\pi \omega^{2}}\int_{\Omega}^{\infty}\mbox{d}\omega_{1}\,\omega_{1}\,\mbox{Im}[{\bfsfG}({\bf r}_{\rm a},{\bf r}_{\rm d},\omega_{1})].
\end{equation}
%
This leads to a high-frequency approximation for the squared F{\"o}rster amplitude $w_{\rm F}^{\rm (HF)}(\omega) = (2\pi/\hbar^{2})(\omega/(\varepsilon_{0} c^{2}))^{2}|{\bm  \mu}_{\rm a}^{*}\cdot {\bfsfG}_{\rm S}^{\rm (HF)}({\bf r}_{\rm a},{\bf r}_{\rm d},\omega)\cdot {\bm  \mu}_{\rm d}|^2$, and a  high-frequency approximation the FRET rate $\gamma_{\rm F}^{\rm (HF)}$:
%
\begin{equation}\label{eq:FRETrateHF}
\gamma_{\rm F}^{\rm (HF)} = \int_{-\infty}^{\infty} \mbox{d}\omega\,\sigma_{\rm a}(\omega)\, w_{\rm F}^{\rm (HF)}({\bf r}_{\rm a},{\bf r}_{\rm d},\omega)\, \sigma_{\rm d}(\omega).
\end{equation}
%
In Figure~\ref{Fig:cut-off_freq}(c) the two approximated FRET rates $\gamma_{\rm F}^{\rm (L)}$ and $\gamma_{\rm F}^{\rm (HF)}$ are compared for the ideal mirror, both scaled by the exact FRET rate $\gamma_{\rm F}$, as a function of the bandwidth $\Omega$.
The donor-acceptor distance and the distance to the mirror are fixed.
Indeed $\gamma_{\rm F}^{\rm (L)}$ is the more accurate approximation of the two, yet $\gamma_{\rm F}^{\rm (HF)}$ is not a bad approximation at all: by only integrating in Eq.~(\ref{newformula_HF}) over high frequencies $\omega_{1}  \ge \Omega = 10 \omega_{\rm d}$, $\gamma_{\rm F}^{\rm (HF)}$ is accurate to within about $7 \%$.
If we take a narrower -- yet still broad -- frequency bandwidth, for example up to $\Omega = 2 \omega_{\rm d}$ (in the UV), we still neglect the LDOS in the whole visible range.
Nevertheless Figure~\ref{Fig:cut-off_freq}(c) shows that for $\Omega = 2 \omega_{\rm d}$ the two approximations $\gamma_{\rm F}^{\rm (L)}$ and $\gamma_{\rm F}^{\rm (HF)}$ agree to a high accuracy with the exact rate $\gamma_{\rm F}$.
Therefore, Figures~\ref{Fig:cut-off_freq} and \ref{Fig:Taylor_vs_distance} show that for the ideal mirror there is essentially no dependence of the FRET rate on the frequency-integrated LDOS at visible frequencies, and only a weak dependence on the frequency-integrated LDOS at UV frequencies and beyond.
We note that this conclusion is different from and complementary to the one in Sec.~\ref{Sec:Foerster_versus_total}, where the FRET rate was found to not depend on the LDOS at one frequency, namely {\em at the transferred energy $\hbar \omega_{\rm d}$}.

\section{Discussion}
In this section, we discuss consequences of our theoretical results to experiments, first regarding relevant length scales.
We have performed analytical calculations and plotted rates versus reduced lengths, namely the reduced distance to the mirror $z \omega / c = 2 \pi z / \lambda$, and the reduced donor-acceptor distance $r_{\rm da} / \lambda$.
To increase the relevance of our results to experiments and applications, we have plotted in several figures additional abscissae for absolute length scales that pertain to a particular choice of the donor emission wavelength $\lambda_{d}$.
Here we have chosen $\lambda_{d} = 2\pi/\omega_{d} = (2\pi \cdot 100) {\rm nm} \simeq 628~{\rm nm}$, a figure that we refer to as a "Mermin-wavelength"~\cite{footnote_Mermin}, as it simplifies the conversion between reduced units and real units to a mere multiplication by $100 \times$.

Figures~\ref{Fig:FRET_dist} and~\ref{Fig:Char_Distances} characterize the distance dependence to the mirror.
The range where both the total and the FRET rate are controlled by the distance to the mirror is in the range $z < 20~{\rm nm}$.
This range is set by the donor-acceptor distance that is for most typical FRET pairs in the order of $r_{\rm da} = 10 {\rm nm}$, in view of typical F{\"o}rster distances of the same magnitude~\cite{Lakowicz2006}.
Interestingly, while the energy transfer is in this range ($z < r_{\rm da}$) not controlled by the LDOS, the transfer rate itself is nevertheless controlled by precise positioning near a mirror.
An example of a method that could be used to achieve such control at optical wavelengths is by attaching emitters - such as molecules or quantum dots - to the ends of brush polymers with sub-10 nm lengths~\cite{Schulz2014Chains}.
With Rydberg atoms, it appears to be feasible to realize the situation $z < r_{\rm da}$, albeit in the GHz frequency range~\cite{Ravets2014NP}.

Figure~\ref{Fig:interpret_vs_DA-distance} characterizes the donor-acceptor distance dependence of the transfer rate.
It is apparent that F{\"o}rster transfer dominates in the range
$r_{\rm da} < 20 {\rm nm}$, a length scale much smaller than the wavelength of light.
In the range $r_{\rm da} > 100$~nm, energy transfer is dominated by radiative transfer, which is reasonable as this distance range becomes of the order of the wavelength.

Let us now briefly discuss broadband LDOS control.
If one insists on invoking the LDOS to control the FRET rate, Figure~\ref{Fig:cut-off_freq} shows that one must control the LDOS over a huge bandwidth that ranges all the way from zero frequency ('DC') to a frequency $\Omega$ that is on the order of $10$ times the donor emission frequency $\omega_{d}$.
This agrees with the qualitative statements in Ref.~\cite{Dood2005PRB}.
If we consider the Mermin-wavelength $628~{\rm nm}$, the upper bound on the LDOS bandwidth corresponds to a wavelength of $62.8 {\rm nm}$~nm, which is deep in the vacuum ultraviolet (VUV) range.
At these very short wavelengths, all materials that are commonly used in nanophotonic control - be it dielectrics such as silica, semiconductors such as silica, or metals such as silver or gold - are strongly absorbing.
In practice, the optical properties of common nanophotonic materials deviate from their commonly used properties at wavelengths below $200$ to $250~{\rm nm}$, which corresponds to $\Omega < 3 \omega_{d}$.
Yet, even if one were able to control the LDOS over a phenomenally broad bandwidth $0 < \Omega < 3 \omega_{d}$,  Figure~\ref{Fig:Taylor_vs_distance} shows that the broadband LDOS-integral contributes negligibly - much less than $10^{-3}$ - to the F{\"o}rster transfer rate.
Thus, with the current state of the art in nanofabrication, true LDOS-control of F{\"o}rster energy transfer seems to be extremely challenging.

The importance of distinguishing FRET from other energy-transfer mechanisms has also been emphasized by Govorov and co-workers~\cite{Govorov:2007a}, who studied plasmon-enhanced F{\" o}rster transfer near conducting surfaces.
They predict that near metal surfaces the FRET rate can be enhanced by much more than the factor 4 reported here for an ideal mirror, but that a strong enhancement occurs only near the plasmon peak, \emph{i.e.}, in a highly dispersive region, while on the other hand not too close to the resonance since otherwise loss makes F{\"o}rster transfer invisible. Analogous FRET enhancements only in the highly dispersive region near a resonance were found in Ref.~\cite{Dung2002PRA} for single-resonance Drude-Lorentz type media.
In Ref.~\cite{Velizhanin:2012a}, FRET near graphene is also clearly distinguished from long-range plasmon-assisted energy transfer.
As a future extension of our work, it will be interesting to study plasmon-enhanced FRET and long-range energy transfer rates by considering a mirror with a resonance.

How do our theoretical results compare to experiments?
Our theoretical findings support the FRET-rate and spontaneous-emission rate measurements by Blum et al.~\cite{Blum2012PRL}, where it was found that F{\"o}rster transfer rates are unaffected by the LDOS.
Our findings also agree with the results of Refs.~\cite{Dood2005PRB,Rabouw2014NC,Tumkur2014FD,Konrad2015NS}.
What about other experimental studies that do report a relation between FRET rate and LDOS?
Assuming our theory to be correct, this discrepancy can mean three things.
{\em First}, it could be that in those experiments the energy transfer  between donor and acceptor separated by a few nanometers was not dominated by F{\" o}rster transfer. This might occur for energy transfer within a high-Q cavity, but otherwise does not seem to be a probable explanation.
{\em Second}, our theory leaves open the possibility that there is a correlation between FRET rates and LDOS in case of strong dispersion and/or loss in the frequency overlap range of donor and acceptor spectra, in case of plasmon-mediated FRET for example, because in that case our theory does not apply. 
However, if the correlation between LDOS and FRET rates indeed relies on strong dispersion, then this correlation would be a particular relation rather than the sought general relation. Moreover, Ref.~\onlinecite{GonzagaGaleana:2013a} theoretically studies spontaneous-emission and FRET rates near metal surfaces  but does not report a general linear relation between them.
Technically, in case of non-negligible absorption and concomitant complex dielectric function, the concept of the LDOS breaks down, but FRET rates can still be compared to the imaginary part of the Green function.
As a {\em third} possible reason why our theory does not predict a relation between FRET rates and LDOS while some experiments do, we should mention that our theory does not include typical aspects of experiments, such as incompletely paired donors, cross-talk between dense donor-acceptor pairs, inhomogeneously distributed donor-acceptor distances, or transfer rates influenced by strongly inhomogeneous field distributions that may occur near nanoparticles or slits and indentations.
While describing these effects adds substantial complexity to our simple model, they may be taken into account by judicious choices of simple scatterers such as spheres or dipoles.
Nevertheless, it does not seem likely that such particular additional effects will induce a general dependence of the FRET rate on the LDOS.

Regarding the subject of quantum information processing, FRET is a mechanism by which nearby ($< 10$ nm) qubits may interact~\cite{John1991PRB, Barenco1995PRL, Lovett2003PRB, Reina2004PRL, Nazir:2005a, Unold2005PRL}, intended or not. Lovett \emph{et al.}~\cite{Lovett2003PRB} considered the implications of F{\"o}rster resonance energy transfer between two quantum dots.
In one implementation, it was found that it is desirable to
suppress the F{\"o}rster interaction in order to create entanglement using biexcitons.
In another implementation, it was found that F{\"o}rster resonance energy transfer should not be suppressed, but rather switched.
There is a growing interest in manipulating the LDOS, either suppressing it by means of a complete 3D photonic band gap~\cite{Leistikow2011PRL}, or by ultrafast switching in the time-domain~\cite{Thyrrestrup2013OE}. It follows from our present results that these tools cannot be used to also switch or suppress F{\"o}rster resonance energy transfer between quantum bits.
On the other hand, our results do indicate that FRET-related quantum information processing may be controlled by carefully positioning the interacting quantum systems (\emph{i.e.}, the quantum dots) in engineered inhomogeneous dielectric environments.

\section{Conclusions}
Using an exactly solvable analytical model, we have seen that F{\"o}rster resonance energy transfer rate from a donor to an acceptor differs from the one in a homogeneous medium in close vicinity of an ideal mirror.
For two particular dipole configurations, we found that the FRET transfer rate is inhibited (to $0$) or markedly enhanced ($4 \times$).
Thus, even this simple model system offers the opportunity to control energy transfer rates.
It turns out that differences in FRET rates as compared to those in a homogeneous medium are only noticeable at distances to the mirror on the order of the donor-acceptor distance $r_{\rm da}$ or smaller.
This distance together with the wavelength are the only natural length scales in this simple problem.
Since $r_{\rm da}$ is typically less than $10~{\rm nm}$, that is, orders of magnitude smaller than an optical wavelength, any substantial variations in the FRET rates due to the nanophotonic environment occur on a distance scale on which the LDOS does not vary appreciably in the dielectric medium near the ideal mirrror.

On the larger distance scale of an optical wavelength away from the mirror, there are well-known LDOS variations.
At these larger distances, the FRET rate is constant and the same as in a homogeneous medium.
So the one quantity varies appreciably when the other does not, and \emph{vice versa}.
Therefore, we conclude that the FRET rate does not correlate with the partial or total LDOS.
This particular example already suffices to conclude that {\em in general}, the FRET rate does not correlate - neither linearly, nor quadratically, or otherwise - with the LDOS.

How large is the class of environments for which FRET rate and LDOS are uncorrelated?
We have derived as one of our main results the simple and exact expression~(Eq.~\ref{eq:FRETrate2}) for the FRET rate in an inhomogeneous medium.
As a consequence, it follows that F{\"o}rster energy transfer rates are independent of the LDOS at the transferred photon frequency in all nanophotonic media where material dispersion and loss can be neglected in the donor-acceptor frequency overlap interval.
Other main results are the exact expression~(Eq.~\ref{newformula}) of the static Green function in terms of a frequency integral over the imaginary part of the total Green function, and the corresponding approximate relation~(Eq.~\ref{eq:FRETrateLDOS}) between FRET rate and the frequency-integrated LDOS.
We used the latter relation to show that FRET rates near an ideal mirror are numerically independent of the LDOS, even when integrating the LDOS over all visible frequencies.
We have also argued why the same will be true for other media with weak material dispersion as well.

We have emphasized that not all energy transfer is F{\"o}rster  energy transfer, and that for typical F{\"o}rster donor-acceptor distances below $10~{\rm nm}$, the energy transfer is dominated by F{\"o}rster transfer, as the example of the ideal mirror has also shown.
For arbitrary photonic environments we defined F{\"o}rster transfer as being mediated by virtual photon exchange, the strength of which is determined by the static Green tensor, which in homogeneous media gives rise to the characteristic $1 / (n^{4}r_{\rm da}^{6})$-dependence of the F{\"o}rster resonance energy transfer rate.

\section*{Acknowledgments}
It is a pleasure to thank Bill Barnes, Christian Blum, Ad Lagendijk, Asger Mortensen, and Allard Mosk for stimulating discussions, and Bill Barnes for pointing out Ref.~\cite{footnote_Mermin}.
MW gratefully acknowledges support from the Villum Foundation via the VKR Centre of Excellence NATEC-II and from the Danish Council for Independent Research (FNU 1323-00087).
The Center for Nanostructured Graphene is sponsored by the Danish National Research Foundation, Project DNRF58.
WLV gratefully acknowledges support from FOM, NWO, STW, and the Applied Nanophotonics (ANP) section of the MESA+ Institute.
\appendix

\section{Scaling with donor-acceptor distance of F{\"o}rster transfer rate}\label{App:Scaling}
Here we show that the homogeneous-medium  F{\"o}rster transfer rate, scaling as $\propto 1/({n}_{\rm h}^{4}r_{\rm da}^{6})$, is an important limiting case also for inhomogeneous media.
Let us assume that the donor and acceptor are separated by a few nanometers, experiencing the same dielectric material with a dielectric function $\varepsilon_{\rm h}$, within an inhomogeneous nanophotonic environment.
In all of space, we define the optical potential $\bfsfV({\bf r},\omega) = -[ \varepsilon({\bf r}) - \varepsilon_{\rm h}](\omega/c)^{2}\bfsfI$, so that the optical potential vanishes in the vicinity of the donor-acceptor pair.
Then the Green function of the medium can be expressed in terms of the homogeneous-medium Green function and the optical potential as
\begin{eqnarray}
&&{\bfsfG}({\bf r}_{\rm a},{\bf r}_{\rm d},\omega)  =  {\bfsfG}_{\rm h}({\bf r}_{\rm a}-{\bf r}_{\rm d},\omega)  \\
&&+ \int\mbox{d}{\bf r}_{1}\,{\bfsfG}_{\rm h}({\bf r}_{\rm a}-{\bf r}_{1},\omega)\cdot {\bf V}({\bf r}_{1},\omega) \cdot {\bfsfG}({\bf r}_{1},{\bf r}_{\rm d},\omega), \nonumber
\end{eqnarray}
which is the Dyson-Schwinger equation for the Green function that controls the energy transfer.
The equation can be formally solved in terms of the T-matrix of the medium as
\begin{eqnarray}\label{GasT}
&&{\bfsfG}({\bf r}_{\rm a},{\bf r}_{\rm d})  =  {\bfsfG}_{\rm h}({\bf r}_{\rm a}-{\bf r}_{\rm d})  \\
&&+ \int\mbox{d}{\bf r}_{1}\mbox{d}{\bf r}_{2}\,{\bfsfG}_{\rm h}({\bf r}_{\rm a}-{\bf r}_{1})\cdot {\bf T}({\bf r}_{1},{\bf r}_{2}) \cdot {\bfsfG}_{\rm h}({\bf r}_{2}-{\bf r}_{\bf D}), \nonumber
\end{eqnarray}
where the frequency dependence was dropped for readability. The important property of the T-matrix ${\bf T}({\bf r}_{1},{\bf r}_{2},\omega)$ is now that it is only non-vanishing where both ${\bf V}({\bf r}_{1})$ and ${\bf V}({\bf r}_{2})$ are nonzero, so that it vanishes in the vicinity of the donor-acceptor pair. Thus the Green function that controls the energy transfer is given by the sum of a homogeneous-medium Green function and a scattering term. The former is a function of the distance between donor and acceptor, whereas the latter does not depend on the D-A distance, but rather on the distance of donor and acceptor to points in space where the optical potential is non-vanishing.

As the donor-acceptor distance $r_{\rm da}$ is decreased, the homogeneous-medium contribution in Eq.~(\ref{GasT}) grows rapidly, essentially becoming equal to ${\bfsfG}_{\rm h, S}({\bf r}_{\rm a}-{\bf r}_{\rm d},\omega)$ of Eq.~(\ref{eq:Gshom}), whereas the contribution of the scattering term does not change much. So in the limit of  very small $r_{\rm da}$, or when making the distance to interfaces larger, the homogeneous-medium term always wins, and one would find the well-known F{\"o}rster transfer rate of the infinite homogeneous medium $\propto 1/({n}_{\rm h}^{4}r_{\rm da}^{6})$.


\section{Tests of identity~(\ref{newformula})}\label{App:two_tests}
\subsection{Test for a homogeneous medium}\label{App:test_for_homogeneous_media}

The Green function $\bfsfG_{\rm h}({\bf r},\omega)$ for homogeneous media is given in Eq.~(\ref{Ghom_realspace}), and its static part by Eq.~(\ref{eq:Gshom}). The identity~(\ref{newformula}) that relates them can be shown to hold as a tensorial identity;  here we derive the identity for its projection ${\bm \mu}\cdot\bfsfG_{\rm h}({\bf r},\omega)\cdot{\bm \mu}$, where we assume ${\bm \mu}$ to be perpendicular to  ${\bf r}$.
(Physically, this corresponds to energy transfer between donor and acceptor with equal dipoles both pointing perpendicular to their position difference vector.)
The projection of the identity~(\ref{newformula}) that we are to derive has the form
\begin{eqnarray}
&& \frac{\mu^{2}c^{2}}{4 \pi n^{2} \omega^{2}}\frac{1}{r_{\rm da}^{3}}  =  \\
&&\frac{\mu^{2}}{2 \pi^{2} n^{2} \omega^{2}} \frac{1}{r_{\rm da}}
\int_{0}^{\infty} \mbox{d}\omega_{1} \omega_{1} \mbox{Im}\left[ \,e^{ i n \omega r_{\rm da}/c} P(i n \omega r_{\rm da}/c)\,\right]. \nonumber
\end{eqnarray}
Now by integration variable transformation the right-hand side of this equation can be worked out to give
\begin{equation}\label{eq:towards_identity_for_free_space}
\frac{\mu^{2} c^{2}}{2 \pi^{2} n^{2}\omega^{2} r_{\rm da}^{3}}\int_{0}^{\infty}\mbox{d}x \left[ \cos( k x) + x \sin(k x) - \frac{\sin(x)}{x} \right],
\end{equation}
with dummy variable $k$ equal to unity. Now the first two terms within the square brackets do not contribute to the integral since $\int_{0}^{\infty}\mbox{d}x \cos(k x) = \pi \delta(k)$ and $\int_{0}^{\infty}\mbox{d}x x \sin(k x) = -\pi \frac{d}{d k}\delta(k)$, while the third term in the square brackets of Eq.~(\ref{eq:towards_identity_for_free_space}) does contribute since $\int_{0}^{\infty}\mbox{d}x  \sin(x)/x = \pi/2$.
Thus the projection of the identity~(\ref{newformula}) indeed holds for spatially homogeneous media.


\subsection{Test for an ideal mirror}\label{App:test_for_ideal_mirror}
For the ideal mirror we again only consider a projection of the identity~(\ref{newformula}), first projecting onto dipoles corresponding to the parallel configuration of Fig.~\ref{Fig:situation}. The Green function for the ideal mirror is given in Eq.~(\ref{eq:G_for_ideal_mirror}), and its static part for the parallel configuration by Eq.~(\ref{mu_Gs_mu_parallel}).
Now for this parallel configuration, the projected Green tensor consists of a homogeneous-medium and a reflected part, and so does the projected static Green function. In Sec.~\ref{App:test_for_homogeneous_media} above we already showed that the sought identity indeed holds for homogeneous media. So the remaining task is to show that the identity~(\ref{newformula}) holds separately for the reflected parts of the projected Green functions. This is not difficult since mathematically the frequency integral that is to be performed is the same as for the homogeneous medium; only the distance parameter $r_{\rm da}$ is to be replaced by $\sqrt{r_{\rm da}^{2} + 4 z^{2}}$. Thus the projection of the identity~(\ref{newformula}) onto the parallel dipole directions indeed holds. The qualitative novelty as compared to the homogeneous-medium case is that we thus show that the identity holds irrespective of the distance $z$ of the FRET pair to the mirror. We also checked (not shown) that the identity~(\ref{newformula}) holds for the projection onto  perpendicular dipoles, i.e. as in the perpendicular configuration of Fig.~\ref{Fig:situation}.


\section{Accuracy of the approximate expressions~(\ref{newformula2}) and (\ref{eq:FRETrateHF})}\label{App:accuracy_tests_of_LDOS_approximation}
To test the accuracy of the LDOS approximation ${\bfsfG}_{\rm S}^{\rm (L)}({\bf r}_{\rm a},{\bf r}_{\rm d},\omega)$ of the static Green function, it is convenient to use Eq.~(\ref{newformula}) to rewrite Eq.~(\ref{newformula2}) as
\begin{eqnarray}\label{newformula2App}
&&{\bfsfG}_{\rm S}^{\rm (L)}({\bf r}_{\rm a},{\bf r}_{\rm d},\omega) =
 {\bfsfG}_{\rm S}({\bf r}_{\rm a},{\bf r}_{\rm d},\omega)  \\
&&+\frac{2}{\pi \omega^{2}}\int_{0}^{\Omega}\mbox{d}\omega_{1}\,\omega_{1}\,\mbox{Im}[{\bfsfG}({\bf r}_{\rm d},{\bf r}_{\rm d},\omega_{1})-{\bfsfG}({\bf r}_{\rm a},{\bf r}_{\rm d},\omega_{1})], \nonumber
\end{eqnarray}
In this form, the approximate static Green function is equal to the exact expression plus an integral over a finite interval of a well-behaved integrand. Likewise, to test the accuracy of the high-frequency approximation ${\bfsfG}_{\rm S}^{\rm (HF)}({\bf r}_{\rm a},{\bf r}_{\rm d},\omega)$ defined in Eq.~(\ref{eq:FRETrateHF}) of the static Green function, it is useful to rewrite it as
\begin{eqnarray}\label{eq:FRETrateHFApp}
{\bfsfG}_{\rm S}^{\rm (HF)}({\bf r}_{\rm a},{\bf r}_{\rm d},\omega) & = & {\bfsfG}_{\rm S}({\bf r}_{\rm a},{\bf r}_{\rm d},\omega)  \\
&-&\frac{2}{\pi \omega^{2}}\int_{0}^{\Omega}\mbox{d}\omega_{1}\,\omega_{1}\,\mbox{Im}[{\bfsfG}({\bf r}_{\rm a},{\bf r}_{\rm d},\omega_{1})], \nonumber
\end{eqnarray}
Again the integrand is well-defined, \emph{i.e.} non-diverging, over the entire finite integration interval.


\subsection{Accuracy of LDOS approximation for homogeneous media}\label{App:accuracy_for_homogeneous_medium}
We estimate the accuracy of Eq.~(\ref{newformula2App}) for the Green function~(\ref{Ghom_realspace}) of a homogeneous medium. By taking the projection onto dipole vectors both on the left and right, we find
\begin{eqnarray}\label{projected_G_integral}
&& \mbox{Im}\int_{0}^{\Omega}\mbox{d}\omega_{1}\,\omega_{1}\,{\bm \mu}\cdot[{\bfsfG}_{\rm h}({\bf r}_{\rm d},{\bf r}_{\rm d},\omega_{1})-{\bfsfG}_{\rm h}({\bf r}_{\rm a},{\bf r}_{\rm d},\omega_{1})]\cdot{\bm \mu} = \nonumber \\
&& - \frac{\mu^{2}n}{4 \pi c}\int_{0}^{\Omega}\mbox{d}\omega_{1} \omega_{1}^{2} \left[ \frac{2}{3} - h(D)\right],
\end{eqnarray}
where $D= n \omega d/c$ and for convenience we defined the function $h(x) \equiv \cos(x)/x^{2} + \sin(x)(1 - 1/x^{2})/x$.
So here (and also for the mirror below) we must determine integrals of the type
\begin{eqnarray}\label{H_integral_identity}
H(\Omega, a) & = & \int_{0}^{\Omega}\mbox{d}\omega_{1} \omega_{1}^{2} h(\omega_{1} a)   \\
& = &  (\Omega/A)^{3}\left[ \sin(A) - A\cos(A) - Si(A)\right].  \nonumber
\end{eqnarray}
where $A = \Omega a$ and $Si[x] = \int_{0}^{x}\mbox{d}t \sin(t)/t$ is the sine integral.
For $\Omega a \ll 1$ we find the approximation
\begin{equation}
H(\Omega, a)  = \frac{2}{9}\Omega^{3} - \frac{2}{75}\frac{(\Omega a)^{5}}{a^{3}}.
\end{equation}
With this result, we find that the relative error of making the LDOS approximation ${\bfsfG}_{\rm h, S}^{\rm (L)}({\bf r}_{\rm a},{\bf r}_{\rm d},\omega)$ of Eq.~(\ref{newformula2App}) for the Green function ${\bfsfG}_{\rm h, S}({\bf r}_{\rm a},{\bf r}_{\rm d},\omega)$ is
\begin{equation}
\frac{{\bm \mu}\cdot\left( {\bfsfG}_{\rm h, S}^{\rm (L)} - {\bfsfG}_{\rm h, S}\right)\cdot {\bm \mu}}
{{\bm \mu}\cdot{\bfsfG}_{\rm h, S}\cdot {\bm \mu}} = - \frac{4}{75 \pi} (\Omega r_{\rm da}n/c)^{5}.
\end{equation}
This fifth-power dependence shows that for homogeneous media the LDOS approximation is excellent as long as $\Omega r_{\rm da}n/c \ll 1$, which for typical F{\"or}ster distances of a few nanometers corresponds to a frequency bandwidth $\Omega$ of order $10 \omega_{\rm d}$ in which the LDOS approximation can be made, where $\omega_{\rm d}$ is a typical optical frequency (e.g., the donor emission frequency).


\subsection{Accuracy of LDOS approximation for the ideal mirror}\label{App:accuracy_for_mirror}
For the {\em parallel configuration} near the ideal mirror, we find  Eq.~(\ref{projected_G_integral}), but with the integrand on the right-hand side replaced by
\begin{equation}
- \frac{\mu^{2}n}{4 \pi c}\omega_{1}^{2}\left\{ \left[ \frac{2}{3} - h(D_{1})\right] - \left[ h(2 Z_{1}) - h(U_{1}) \right]\right\},
\end{equation}
where $D_{1} = \omega_{1}r_{\rm da}n/c$, $Z_{1} = \omega_{1} z n/c$, and $U_{1} = \sqrt{D_{1}^{2} + 4 Z_{1}^{2}}$.
So we can identify both a homogeneous-medium and a scattering contribution between the curly brackets.
By threefold use of the identity~(\ref{H_integral_identity}) it then follows that ${\bm \mu}\cdot(\bfsfG_{\rm S}^{\rm (L)} - \bfsfG_{\rm S})\cdot{\bm \mu}$ equals
\begin{equation}\label{parallel_approx_result}
- \frac{\mu^{2}n}{2 \pi^{2}\omega^{2}c}\left\{ \left[\frac{2}{9}\Omega^{3} - H(\Omega,\frac{n r_{\rm da}}{c})\right] - \left[ H(\Omega,\frac{2 n z}{c}) - H(\Omega,\frac{n u}{c})\right] \right\},
\end{equation}
where $ u =\sqrt{r_{\rm da}^{2} + 4 z^{2}}$.

For the {\em perpendicular configuration} near the ideal mirror, it can be found that the integrand of Eq.~(\ref{projected_G_integral}) is instead replaced by the slightly longer expression
\begin{eqnarray}\label{perp_approx_result}
&& - \frac{\mu^{2}n }{4 \pi c}\omega_{1}^{2}\left\{ \left( \frac{2}{3} - 2 h(2 Z_{1}) + 2 \frac{\sin(2 Z_{1})}{2 Z_{1}}\right) - h(D_{1})  \right.\nonumber \\
&& \left. - h(U_{1}) - \frac{4 z^{2}}{\sqrt{r_{\rm da}^{2} + 4 z^{2}}}\left[2 \frac{\sin(U_{1})}{U_{1}} - 3 h(U_{1}) \right] \right\}.
\end{eqnarray}
The frequency integral can again be performed using the identity~(\ref{H_integral_identity}) and a standard integral of the type $\int \mbox{d}x\, x \sin(x)$.
The resulting expression for ${\bm \mu}\cdot(\bfsfG_{\rm S}^{\rm (L)} - \bfsfG_{\rm S})\cdot{\bm \mu}$, and the corresponding result~(\ref{parallel_approx_result}) for the parallel configuration are both used in Figs.~\ref{Fig:cut-off_freq} and \ref{Fig:Taylor_vs_distance}.


\end{document}